\documentclass{WileyMSP-template}

\usepackage{amsmath,amsfonts}

\usepackage{algorithmic}
\usepackage{algorithm}
\usepackage{array}
\usepackage{textcomp}
\usepackage{stfloats}
\usepackage{url}
\usepackage{bm}
\usepackage{siunitx}
\usepackage{verbatim}
\usepackage{makecell}
\usepackage{multirow}
\usepackage{colortbl}
\usepackage{hyperref}
\usepackage{graphicx}
\usepackage{rotating}
\usepackage{tabularx}
 \usepackage{lineno}
 
\DeclareMathOperator*{\argmax}{arg\,max}

\begin{document}

\pagestyle{fancy}

\title{Social Interaction-Aware Dynamical Models and Decision Making for Autonomous Vehicles}

\maketitle


\author{Luca Crosato\textsuperscript{†}}
\author{Kai Tian\textsuperscript{†}}
\author{Hubert P. H. Shum}
\author{Edmond S. L. Ho}
\author{Yafei Wang}
\author{Chongfeng Wei*}

\dedication{\textsuperscript{†}: these two authors contributed equally to this work as co-first authors. *: Corresponding author.}

\begin{affiliations}
Luca Crosato\\
Address: Northumbria University, Newcastle upon Tyne, NE1 8ST, UK; School of Mechanical and Aerospace, Queen’s University Belfast, BT7 1NN, UK\\
Email Address: luca.crosato@northumbria.ac.uk\\
Dr. Kai Tian\\
Address: Institute for Transport Studies, University of Leeds, LS1 9JT, UK\\
Email Address: tiankai\_1993@hotmail.com\\
Dr. Hubert P. H. Shum\\
Address: Department of Computer Science, Durham University, DH1 3LE, UK\\
Email Address: hubert.shum@durham.ac.uk\\
Dr. Edmond S. L. Ho\\
Address: School of Computing Science, University of Glasgow,  G12 8QQ, UK\\
Email Address: Shu-Lim.Ho@glasgow.ac.uk\\
Dr. Yafei Wang\\
Address: School of Mechanical Engineering, Shanghai Jiao Tong University, 200240, China\\
Email Address: wyfjlu@sjtu.edu.cn\\
Dr. Chongfeng Wei\\
Address: School of Mechanical and Aerospace, Queen’s University Belfast, BT7 1NN, UK\\
Email Address: c.wei@qub.ac.uk

\end{affiliations}



\begin{abstract}
\justify
Interaction-aware Autonomous Driving (IAAD) is a rapidly growing field of research that focuses on the development of autonomous vehicles (AVs) that are capable of interacting safely and efficiently with human road users. This is a challenging task, as it requires the autonomous vehicle to be able to understand and predict the behaviour of human road users. In this literature review, the current state of IAAD research is surveyed in this work. Commencing with an examination of terminology, attention is drawn to challenges and existing models employed for modelling the behaviour of drivers and pedestrians. Next, a comprehensive review is conducted on various techniques proposed for interaction modelling, encompassing cognitive methods, machine learning approaches, and game-theoretic methods. The conclusion is reached through a discussion of potential advantages and risks associated with IAAD, along with the illumination of pivotal research inquiries necessitating future exploration.

\end{abstract}



\section{Introduction}
\justify
In the past few years, there has been an increasing interest in the development of technology for AVs, as the recent advancements in Robotics and Machine Learning have enabled Autonomous Driving (AD) engineers to develop algorithms that could tackle the complexity of the autonomous driving task. AVs have the potential to improve traffic quality, reduce traffic accidents, and improve the quality of time spent while travelling \cite{yurtsever2020survey}. Nowadays, more and more AVs are being deployed into the real world, sharing the environment with other human road users. This has raised concerns that AVs could not be able to understand and interact smoothly with other human road users, potentially leading to traffic dilemmas and safety issues \cite{millard2018pedestrians}. In order to operate in an efficient and safe manner, AVs need to behave in a human-like fashion and generate optimal behaviours that take the interactions with other human road users into account \cite{maurer2016autonomous}. This is critical for the reduction of potential traffic conflicts. For example, cautious but unnecessary stops at intersections might cause rear-end accidents.
In order to develop fully automated vehicles, advances in many aspects of AV technology are required, ranging from perception, decision-making, planning, and control \cite{pendleton2017perception, schwarting2018planning}. When it comes to predicting the behaviour of surrounding human road users and taking decisions accordingly for AVs, the interactions with surrounding human road users become increasingly important, as the AV actions affect their behaviour and vice versa \cite{hubmann2017decision}. 

The purpose of this paper is to provide an exhaustive survey of state-of-the-art techniques in interaction-aware motion planning and decision in the context of autonomous driving. In particular, the text first covers human road user \textbf{behavioural models to highlight what factors influence the decisions that human road users make on the road}. Driver and Pedestrian Behavioural Models are relevant to AVs for multiple reasons. Firstly, they can be used to assess and predict what road users that surround the AV will do. Secondly, they can aid in the development of human-like AV behaviour. Therefore, they hold both a predictive value as well as add relevant insights for model/system design. \\
\begin{figure}[t]
    \centering
    \includegraphics[width=1.0\linewidth]{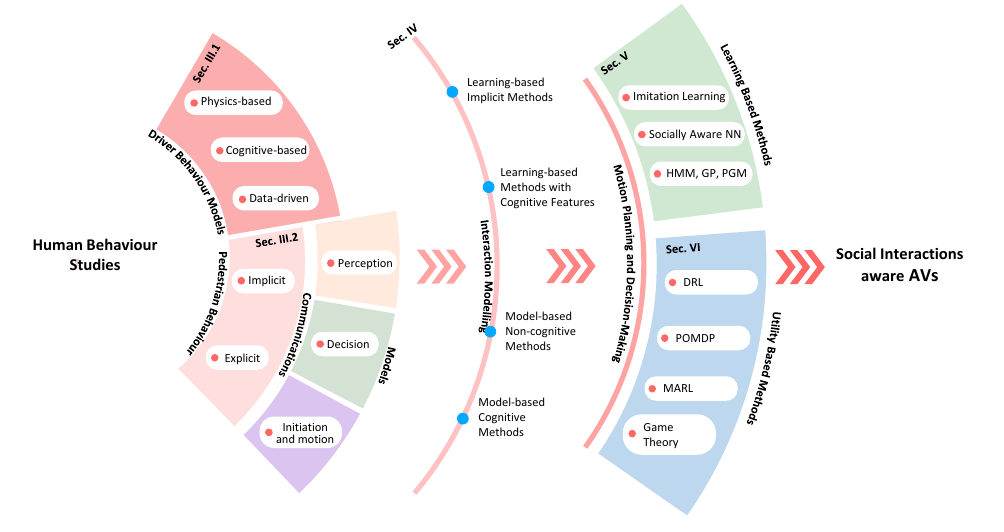}
    \caption{Flow chart: form human behaviour to social interaction aware AVs.}
    \label{fig:flow chart}
\end{figure}

This review consists of 5 main sections which cover different areas in interaction-aware autonomous driving. The terminology used in interaction-aware autonomous driving is introduced in Section \ref{sec:Terminology}.  Please refer to Figure \ref{fig:flow chart} for an overview of the paper structure. 

Section \ref{sec:Human Behaviour Studies} will cover human factors studies on what affects human decision making while driving, as well as pedestrian behavioural studies.
Section \ref{sec:Interaction Modelling} gives a broad overview and classification of existing techniques that are used in interaction modelling. Finally, Sections \ref{sec:Learning Based Methods} and \ref{sec:Utility Based Methods} cover state-of-the-art techniques used for motion-planning and decision-making in interactive scenarios. 

While autonomous driving has been an active research area in recent years, most of it focuses on scenarios involving only vehicles. There is more limited work that addresses heterogeneous scenarios, which include both vehicles and pedestrians. In this paper, the focus is on heterogeneous scenarios, but Sections \ref{sec:Learning Based Methods} and \ref{sec:Utility Based Methods} will also cover related work that deals with scenarios without pedestrians. This is because the techniques used in these papers can be easily adapted to mixed traffic scenarios, or they can offer important insights into how to deal with the general problem of mixed traffic scenarios.

\begin{figure}[t]
    \centering
    \includegraphics[width=1.0\linewidth]{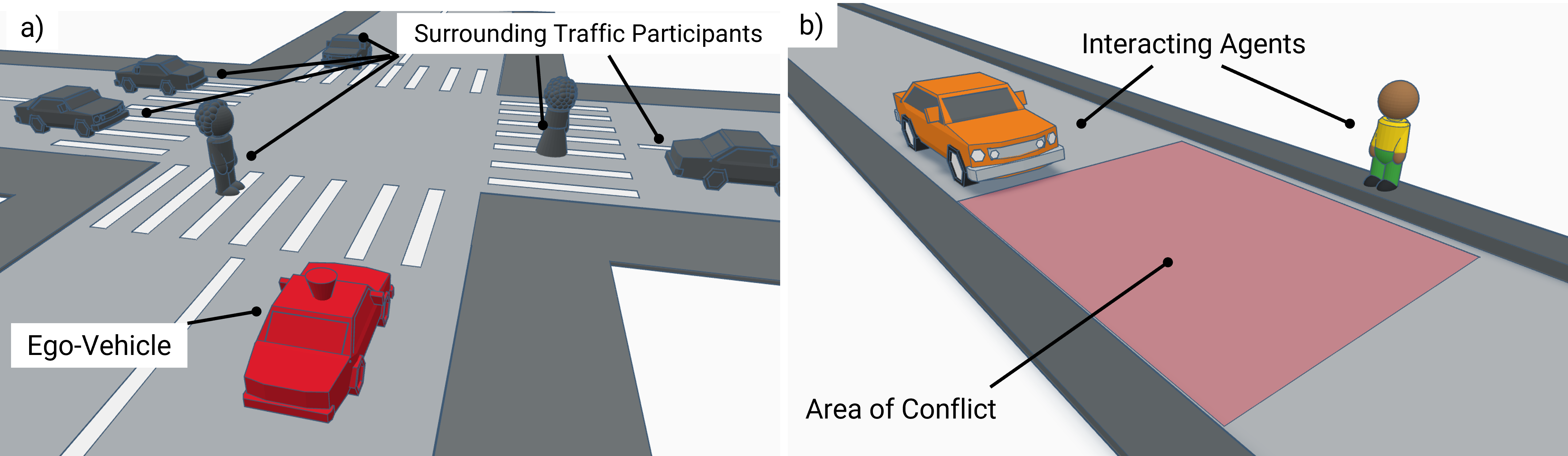}
    \caption{a) the ego-vehicle is controlled by the autonomous system, whereas surrounding traffic participants act on their own will. b) two agents interacting with each other determine an area of conflict. Reproduced with permission.\textsuperscript{[Ref.]} Copyright Year, Publisher.}
    \label{fig:considerations}
\end{figure}

\section{Terminology in Interaction-Aware Autonomous Driving}\label{sec:Terminology}
\justify
Before discussing the recent advances in interaction-aware motion-planning and decision-making, the paper first defines some of the terminology used in this field. In the field of autonomous driving, the term \textit{ego-vehicle} refers to the specific vehicle whose behaviour is to be controlled and studied. All other vehicles, cyclists, pedestrians, etc., that occupy a region of space around the ego-vehicle are treated as interactive obstacles and are referred to as \textit{surrounding traffic participants}, see Figure \ref{fig:considerations}a. Since road traffic is unlikely to become fully automated in the near future, AVs will inevitably operate in environments mixed with \textit{human road users} (HRUs), such as human drivers and pedestrians. Therefore, \textbf{interaction-aware} autonomous driving is a field of research that focuses on developing AVs that can safely and efficiently interact with surrounding HRUs. Traditional autonomous driving approaches often treat surrounding HRUs as dynamic obstacles. However, this is not a realistic approach, as they are constantly changing their behaviour to adapt to the current situation.

\begin{figure}[t]
    \centering
    \includegraphics[width=1.0\linewidth]{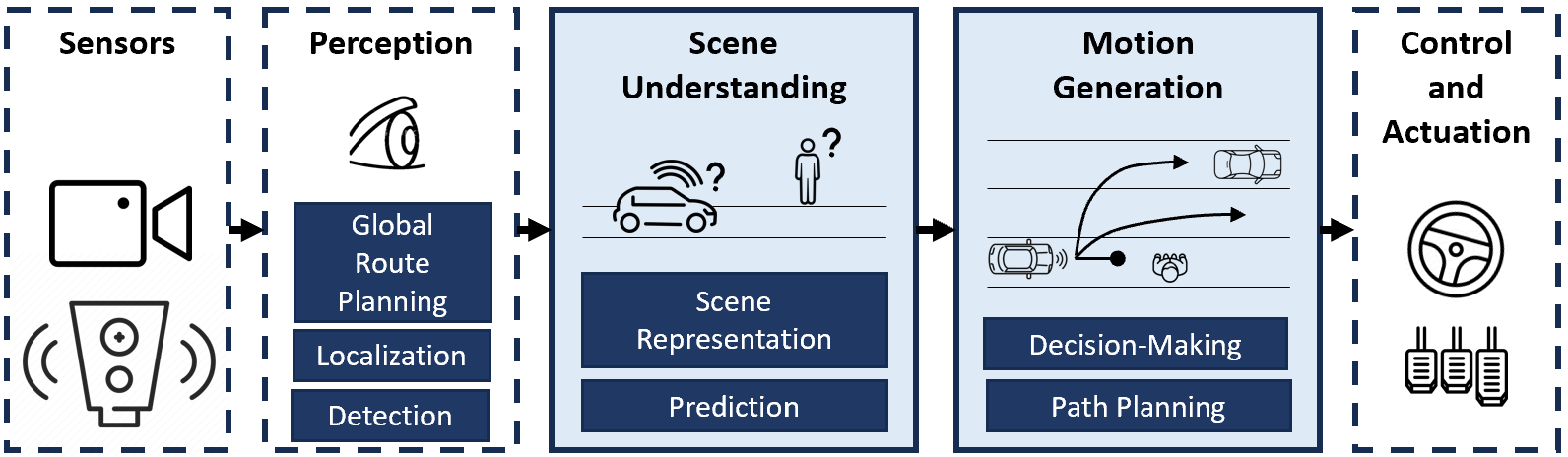}
    \caption{Architecture of AV systems. Solid line boxes identify modules that are closely related to interaction-aware models.}
    \label{fig:interaction_aware_av_pipeline}
\end{figure}

Generally, multiple surrounding HRUs can give rise to \textit{space-sharing conflicts} amongst themselves or with the ego-vehicle: a situation from which it can be reasonably inferred that two or more road users intend to occupy the same region of space at the same time in the near future, see Figure \ref{fig:considerations}b. The agents involved in the conflict are said to display an \textit{interactive behaviour}, which implies that their behaviour would have been different if the space-sharing conflict had not occurred \cite{markkula2020defining}. Moreover, interaction does not necessarily involve conflict. It can be explicit or implicit communications that indicate a road user's intentions and affect HRUs. For instance, drivers could plan their driving strategy based on the turning light signals of vehicles in front, even though the ego-vehicle and vehicles in front are not in the same lane, and there will be no conflict in the near future. Hence, \textit{interactive behaviour} refers to the different courses of action of road users, adapting to the behaviour of others or making requests for reactions and taking actions to achieve their desired goals\cite{swan1988social}. Since interactions happen all the time when driving, it is crucial that the algorithms developed for AVs be aware of the dynamics of the interactions between agents. Such algorithms are said to be interaction-aware and are often the focus of recent autonomous driving research \cite{turnwald2019human}. 

The safe and socially acceptable interaction-aware autonomous driving systems are currently hampered by a number of challenges  \cite{liu2020computing}. One challenge is a lack of innovative theories about how HRUs interact \cite{markkula2023explaining}. This is a difficult task, as the theories to be developed are not limited to predicting and modelling HRUs' behaviour but also exploring behaviour patterns and their underlying mechanisms. Integrating AVs into road traffic as seamlessly as humans would require more advanced behaviour theories and models. Another challenge is the need to develop algorithms that can safely and efficiently interact with other HRUs and produce an AV behaviour that compels human-like standards. Figure \ref{fig:interaction_aware_av_pipeline} shows the main parts that make up an AV system. Raw data from sensors is processed by a Perception Module, which detects the surrounding environment and performs localisation, which allows the generating of a global route plan for the ego-vehicle to reach its target destination. The scene can be further interpreted, and predictions regarding surrounding traffic participants can be performed. Interaction-aware models play a major role in prediction tasks, as agents affect each other's trajectories and decisions. 

Decision-making and path-planning are two of the most important tasks in autonomous driving. They are responsible for determining how the vehicle will move through its environment. Decision-making is the process of choosing an action from a set of possible options. For example, the vehicle may need to decide whether to change lanes, slow down, or stop. Path-planning is the process of generating a safe and feasible trajectory for the vehicle to follow. Decision-making and path-planning are closely related. The decision-making process typically outputs a high-level plan, such as "change lanes to the left." The path-planning process then takes this plan and generates a detailed trajectory that the vehicle can follow. Both tasks must take into account the vehicle's current position, the vehicle's capabilities and the surrounding traffic, which is why interaction-aware models are highly relevant to these two tasks.
From a control system perspective, the dynamics of the vehicle are represented by its states, i.e. position and orientation, and their time derivatives. The state of the environment is determined by the states of all dynamics and static entities. The strictly physical state-space can also be augmented with additional latent-space variables that capture, for example, the intentions \cite{bandyopadhyay2013intention} or the behavioural preferences of surrounding users \cite{schwarting2019social}, which are part of the Scene Understanding system.

\section{Human Behaviour Studies on Interactions} \label{sec:Human Behaviour Studies}
\justify
This section synthesizes empirical and modelling research findings on HRU behaviour, including that of human drivers and pedestrians interacting with AVs or conventional vehicles, especially from a communication perspective. The focus is on research involving road interactions, with the aim of discovering insights that may facilitate the development of interaction-aware AVs. Studies that look at macro-traffic conditions, such as the influences of route choice, weather, or regulation, are beyond this paper's scope.

\subsection{Driver Behaviour Studies}
\justify
Driver behaviour models are used to predict and understand how drivers will behave in different driving scenarios. These models can be used to improve the safety and efficiency of transportation systems and aid in the process of designing AVs. Many different factors can affect driving behaviour, including individual characteristics (age, gender, personality, experience), environmental factors, i.e. road and weather conditions, and social factors, which include the driver's interactions with HRUs \cite{abuali2016driver}. 
A comprehensive overview of Driver Behavioural Models (DBM) in vehicle-vehicle interactions can be found in \cite{negash2023driver}. Here, the focus will be on DBMs that are relevant to vehicle-pedestrian interactions.

\begin{figure}[t]
    \centering
    \includegraphics[width=1.0\linewidth]{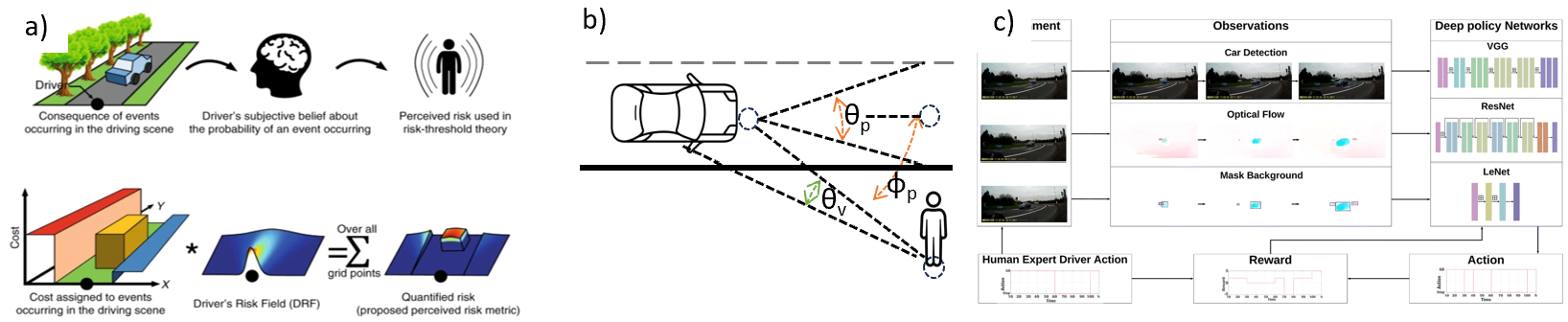}
    \caption{Illustration of Driver Behaviour Models. a) driver risk field from \cite{kolekar2020human}, b) joint theory-based model in \cite{domeyer2022driver}, c) data-driven model in \cite{wang2022imitation}}
    \label{fig:dbm}
\end{figure}

The most common driver behavioural models include: 
\begin{itemize}
    \item \textit{Driver's Risk Field Models}: (Figure \ref{fig:dbm}a) This model predicts how drivers will perceive risk in different driving situations. The DRF model is based on the idea that drivers make decisions based on their perception of risk. The results of \cite{kolekar2020human} suggest that driving behaviour is governed by a cost function that takes into account the effects of noise on human perception and actions. This is similar to how motor-control tasks are governed by cost functions. Risk perception onboard of AVs has also been analysed in \cite{petit2021risk} in driving simulator scenarios.
    \item \textit{Theory Based}: (Figure \ref{fig:dbm}b) perceptual and cognitive models. Models based on perceptual information describe driver's behaviour based on perceptual cues, e.g. distance, vehicle speed, acceleration, expansion angle, reaction times, etc \cite{domeyer2022driver, sun2019interpretable}. 
    Cognitive models outline the internal state flow and motive that regulates the driver's behaviour as a psychological human being \cite{varhelyi1998drivers, markkula2023explaining}.
    \item \textit{Data Driven Models}:  (Figure \ref{fig:dbm}c) this set of methods relies on analysing naturalistic driving data with machine learning to analyse driver behaviour. Data-driven models can learn generative or discriminative \cite{zhou2020driving, shahverdy2020driver, wang2022imitation} models of human behaviour to make predictions about the driver's future decisions or preferred driving style.
    Model validation can be done by comparing predictions with real data and by human-in-the-loop simulations. 
\end{itemize}

Existing research highlights based on naturalistic driving data analyses how drivers behave in the presence of pedestrians. In \cite{rasch2020drivers}, the authors found that drivers tend to maintain smaller minimum lateral clearance and lower overtaking speed when overtaking pedestrians who are walking in the opposite direction, on the lane edge, or when oncoming traffic is present. Minimum lateral clearance and time-to-collision were only weakly correlated with overtaking speed. The results in \cite{sun2022human} show that the vehicle deceleration behaviour is relative to the initial Time To Collision (TTC), subjective judgment of pedestrian crossing intention, vehicle speed, pedestrian position and crossing direction.

There is less attention paid to multi-agent settings where multiple vehicles and pedestrians interact with each other. In \cite{nasernejad2022multiagent}, the authors develop a 
Multi-agent adversarial Inverse Reinforcement Learning (IRL) framework based on data collected at a road intersection to simulate driver and pedestrian behaviour at intersections.

Overall, DBMs are a promising area of research with the potential to significantly improve the safety and efficiency of transportation systems. However, there is still much work to be done in developing and validating these models. Future research should focus on developing more comprehensive models that take into account a wider range of factors, such as the driver's internal state, the environment, and the interactions with other HRUs.

\subsection{Pedestrians Behaviour Studies}
\justify
Since pedestrians are considered the most vulnerable road users, lacking protective equipment and moving more slowly than other road users \cite{el2020pedestrian}, investigating pedestrian behaviour is clearly relevant to the safety and acceptance of AVs interacting with pedestrians. Pedestrian behaviour has been the subject of extensive research for decades \cite{moore1953pedestrian}. The emergence of AVs has recently prompted many new research questions about pedestrian behaviour. Given the large body of work in this area and our aims, this Section examines major studies rather than providing an exhaustive survey. The review covers pedestrian behaviour studies regarding interactions with vehicles from three perspectives: communications, theories and models of crossing behaviour, and AV-involved applications. The aim is to identify and summarize their value for developing interaction-aware AVs.

\subsubsection{Communications}

\begin{figure}[t]
      \centering
     \includegraphics[width=1.0\linewidth]{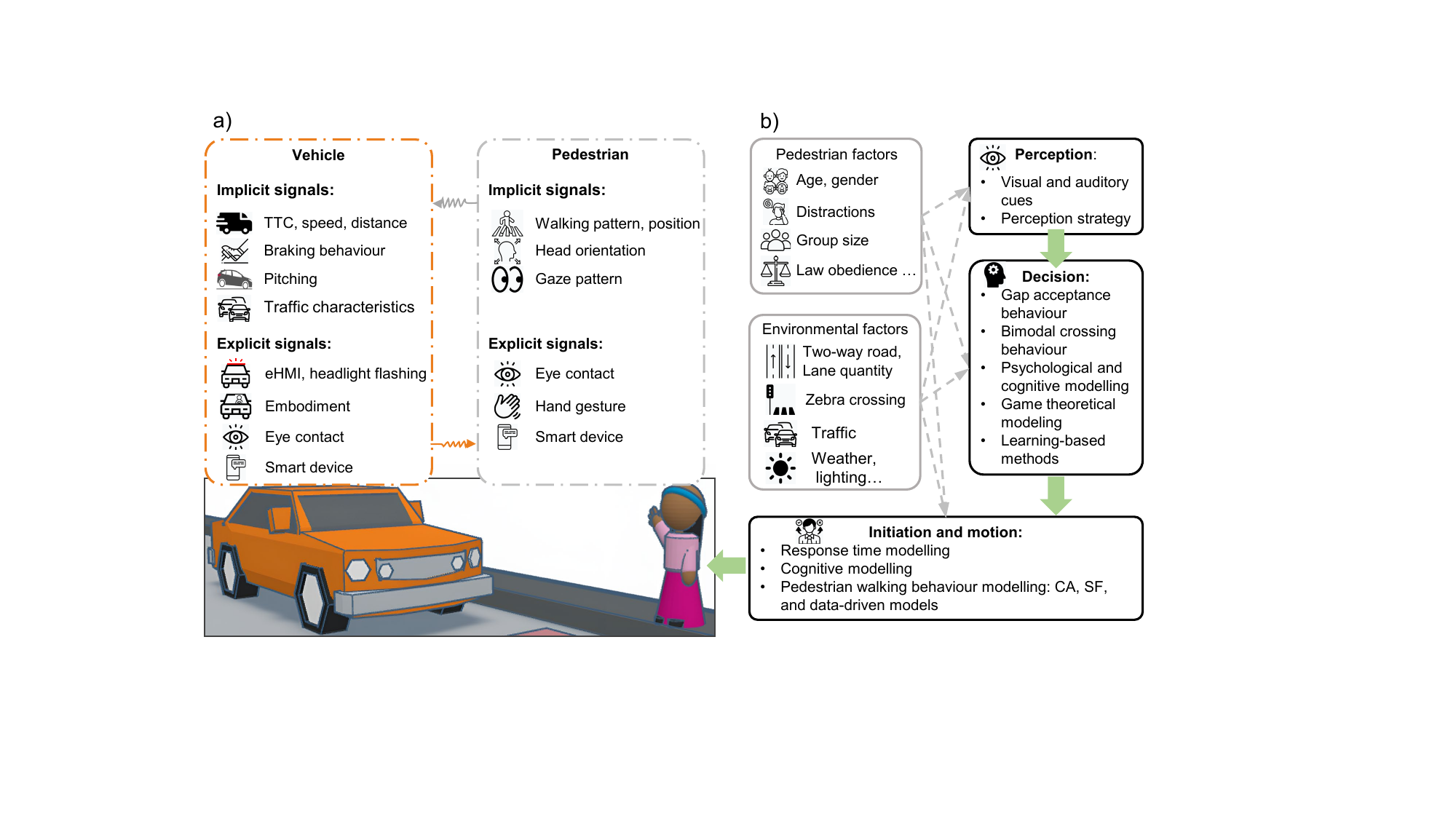}
      \caption{a) Communication between pedestrians and automated vehicles. b) Theories and models for pedestrian crossing perception, decision, initiation, and motion. }
      \label{figlabelkai1}
\end{figure}

In dynamic traffic environments, road users intentionally or unintentionally convey signalling information to one another through their movements and spatial cues, giving rise to both explicit and implicit communication. Research findings concur that AVs' kinematics and signalling information significantly impact pedestrian road behaviour due to the absence of a driver role \cite{markkula2020defining,rasouli2019autonomous,dey2017pedestrian}. Therefore,  the identification of critical motion cues and signals affecting pedestrian road behaviour holds substantial research significance (see Figure \ref{figlabelkai1}a). 

\emph{Implicit communication signals}, such as vehicle kinematic cues, involve road user behaviour that affects its own movement but can be interpreted as cues of intention or movement by another road user \cite{markkula2020defining}. Distance or TTC between approaching vehicles and pedestrians is the most critical Implicit information influencing pedestrian behaviour \cite{lobjois2007age,oxley2005crossing}. Evidence showed that pedestrians tend to rely more on distance than TTC \cite{petzoldt2014relationship}. That is, for the same TTC, more pedestrians crossed when vehicles approached at higher speeds. A recent study showed pedestrians used multiple sources of information from vehicle kinematics instead of relying on one. The impacts of speed, distance, and TTC on pedestrian behaviour were mutual coupling \cite{tian2022explaining}. 

Braking manoeuvre is another critical implicit information influencing pedestrian behaviour in interactions. Vehicle movements correlated with pedestrian trust toward vehicles, emotion, and impact on pedestrian decisions \cite{dey2021communicating,lee2022learning,Tian2023deceleration}. Pedestrians felt comfortable and started crossing quickly when approaching vehicles slowed down early and braked lightly. Harsh braking led to pedestrian avoidance behaviour \cite{dey2021communicating,creech2019pedestrian,risto2017human}. On the other hand, early braking manoeuvres and strong pitching reduced the time required for pedestrians to understand the vehicle's intentions \cite{dietrich2019implicit,ackermannDecelerationParametersTheir2019b}. Vehicles that approached pedestrians slowly while yielding could hinder understanding \cite{dey2021communicating,Tian2023deceleration}.

Traffic characteristics, such as traffic volume and gap sizes, provide implicit information to pedestrians. High traffic volume forced pedestrians to accept small traffic gaps \cite{chandra2014descriptive}, as the increased time cost increased their propensity for risk-taking \cite{zhao2019gap}. However, substantial evidence indicated pedestrians who tended to wait were more cautious and less likely to accept risky gaps \cite{tian2022impacts,lobjois2013effects,yannis2013pedestrian}. The relationship between traffic volume and pedestrian crossing behaviour is context-dependent, potentially influenced by the size and order of gaps in traffic \cite{tian2023deconstructing}.

Furthermore, the pedestrian movement toward the road, presence at the curb, and pedestrian head orientation could convey key implicit information to approaching vehicles \cite{rasouli2017understanding,volz2015feature}. Pedestrians often assert their right of way by stepping onto the road or looking at approaching vehicles \cite{dey2017pedestrian}.  

\emph{Explicit communication signals} involve road user behaviour that conveys signalling information to other road users without affecting one’s own movement or perception \cite{markkula2020defining}. A common case is vehicles transmitting information to pedestrians through an external human-machine interface (eHMI). In the context of AVs, where there is no human driver, eHMIs have gained importance. Substantial evidence supported eHMI’s benefits in pedestrian interactions with AVs \cite{de2019external,dey2021communicating,carmona2021ehmi}. Various types of eHMI prototypes have been proposed, such as headlight \cite{petzoldt2018potential}, light band \cite{lee2022learning}, anthropomorphic symbols \cite{Semcon}, but consensus on the best eHMI form and information to convey remains elusive.

Numerous studies demonstrated that eHMI performance depended on various factors. Pedestrians' familiarity, trust, and interpretation of eHMIs could significantly impact eHMIs' effectiveness in communicating information to pedestrians. For instance, pedestrians better understood conventional eHMIs (flashing headlights) as signals for vehicles yielding than novel eHMIs (light bands) \cite{lee2022learning}. Pedestrians overtrusting eHMIs might lead them to under-rely on vehicle motion cues, which is dangerous if eHMIs fail \cite{kaleefathullah2022external}. Egocentric information transmitted by the eHMI, like ``OK TO CROSS", compelled pedestrians more than allocentric information like ``STOPPING" \cite{bazilinskyycrowdsourced}. Moreover, the reliability of eHMIs was questioned as it might be affected by weather \cite{kooijman2019ehmis}, light condition \cite{rasouli2019autonomous}, and vehicle behaviour \cite{dey2021communicating}. For example, in poor weather, pedestrians could not always read vehicle signs \cite{palmeiro2018interaction}. Pedestrian willingness to cross was unaffected by eHMIs when vehicles did not yield or decelerated aggressively \cite{dey2021communicating,de2019external}. Other concepts, e.g., installing eHMIs on road infrastructures instead of vehicles \cite{farah2022modeling} and combining eHMIs with vehicle motion cues \cite{rettenmaier2021matter,dietrich2019automated}, could outperform pure eHMIs. 

Additionally, from the vehicles' perspective, although less frequent, pedestrians also use explicit signals to communicate with AVs. These signals include eye contact and hand gestures, which are used by pedestrians to ensure they are seen by AVs and to request the right of way  \cite{onkhar2022effect,markkula2020defining,epke2019hand}. To address the absence of a human driver, AVs could utilise a human-like visual embodiment in the driver's seat and wireless communication technology to enhance vehicle-pedestrian communication \cite{rouchitsas2022ghost,furuya2021autonomous,sewalkar2019vehicle}.

\subsubsection{Theories and models of crossing behaviour}
Pedestrian crossing behaviour involves various cognitive processes.  Previous studies \cite{palmeiro2018interaction,coeugnet2019risk,tian2023deconstructing} suggested that there might be three levels of processes involved in constructing pedestrian crossing behaviour in interactions, i.e., perception, decision, initiation and motion. In light of this hypothesis, the following sections synthesise the theories and models of pedestrian crossing behaviour regarding the three cognitive processes (Figure \ref{figlabelkai1}b).
\begin{figure}[t]
      \flushleft
     \includegraphics[width=1.0\linewidth]{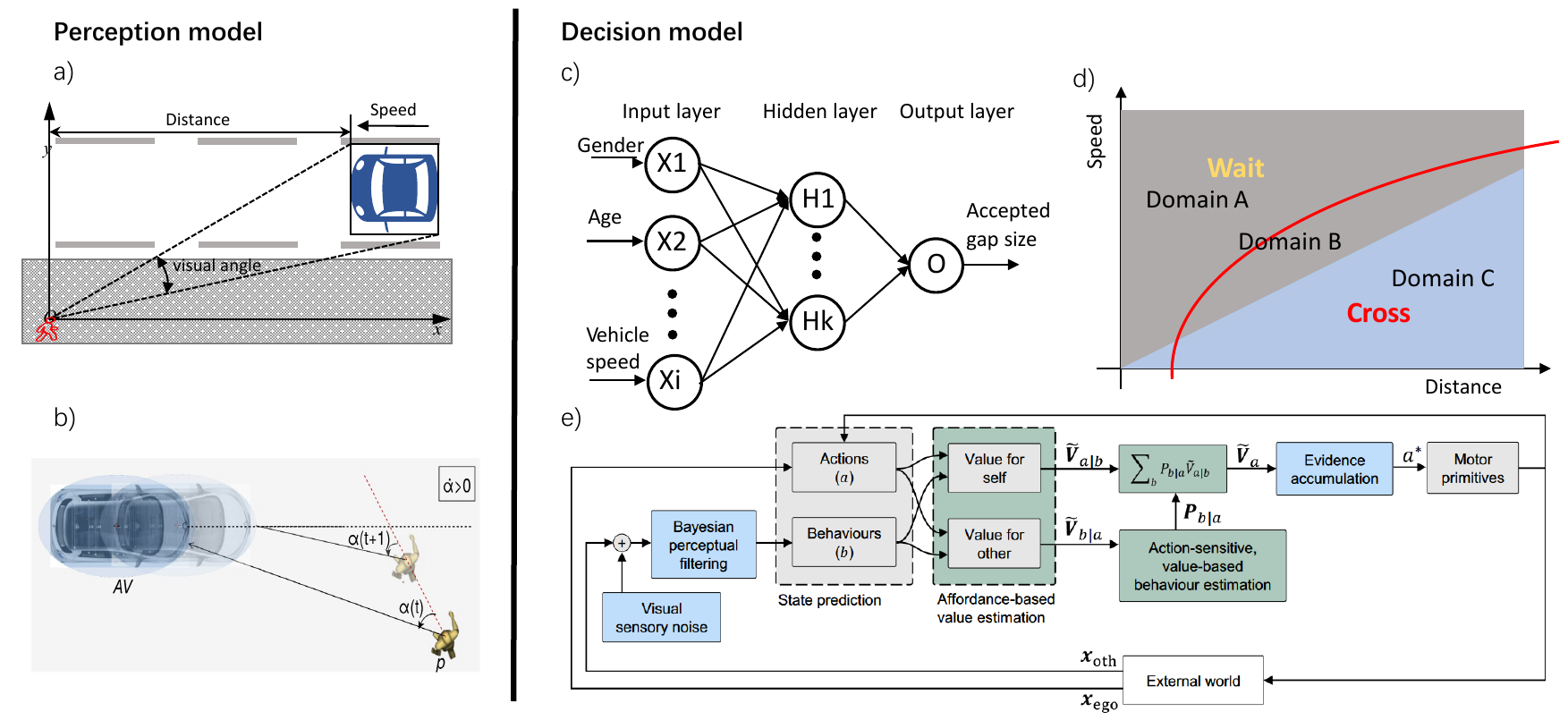}
      \caption{Perception and decision models for pedestrians. a) Visual cues, including $\theta,\dot{\theta},\tau,\dot{\tau}$ \cite{tian2022explaining, tian2023psychological}. b) Bearing angle \cite{predhumeau2022agent}. c) Artificial neural networks \cite{kadali2014evaluation}. d) Speed-distance model \cite{zhu2022defensive}. e) Large computational psychological model \cite{markkula2023explaining}.  }
      \label{figlabelkai2}
\end{figure}
\paragraph{Perception} The visual perception theory, as established by Gibson \cite{gibson2014ecological}, explains that as an object approaches an observer, its image on the retina expands, forming the basis for human collision perception. In crossing scenarios, when the rate of image expansion of a vehicle on the retina reaches a certain threshold, pedestrians perceive that the vehicle is approaching, known as the visual looming phenomenon \cite{hoffmann1994drivers}. A psychophysical model simplifies this expansion rate as the change in the visual angle subtended by the approaching vehicle at the pedestrian's pupil, denoted as  $\dot{\theta}$ (Figure \ref{figlabelkai2}a)\cite{lee1976theory,tian2022explaining}. Recent research suggested that pedestrians use $\dot{\theta}$ as a crucial visual cue to observe approaching vehicles \cite{tian2022explaining, markkula2023explaining}. However, while $\dot{\theta}$ provides spatial information, it does not convey when the vehicle arrives at the pedestrian's position \cite{delucia2008critical}.  In crossing scenarios with yielding vehicles, pedestrians need temporal information to estimate whether the vehicle can stop in time. Lee's mathematical demonstration \cite{lee1976theory} showed that the visual cue $\tau$, representing the ratio of $\theta$ to $\dot{\theta}$, may specify the TTC of the approaching vehicle. Moreover, the first temporal derivative of $\tau$, denoted as $\dot{\tau}$, was relevant for detecting if the current deceleration rate is sufficient to avoid a collision \cite{bardy1997visual}. Furthermore, it was found that pedestrians might visually perceive oncoming collision events under a given angle, i.e., bearing angle, which is the angle between the vehicle and the pedestrian’s line of regard \cite{delucia2015perception} (Figure \ref{figlabelkai2}b).

In addition to visual cues, pedestrian perception may depend on perceptual strategies. Research by Tian et al. \cite{Tian2023deceleration} indicated that pedestrian estimation of vehicle behaviour might be a separate process or a sub-process of crossing decision-making. When there is a large traffic gap, pedestrians tend not to rely on vehicle driving behaviour but rather on gap size. Similarly, Delucia \cite{delucia2008critical} indicated that when collision events are distant, humans tend to use 'heuristic' visual cues, such as $\theta$ and $\dot{\theta}$. However, as collisions become imminent, optical invariants like $\tau$ dominate perception, providing richer spatiotemporal information. 

Besides perception mechanisms, various factors could influence pedestrians' perceptions. Studies showed that elderly or child pedestrians faced a higher collision risk due to age-related perceptual limitations  \cite{lobjois2007age,dommes2021young}. Elderly pedestrians tended to rely more on distance than TTC to judge approaching vehicles, while children struggled to detect vehicles approaching at high speeds \cite{petzoldt2014relationship, wann2011reduced}. Distractions, particularly those involving visual and manual components like smartphone usage, diverted significant attention resources and affected pedestrian observation of traffic conditions \cite{jiang2018effects}. In comparison, cognitive distractions, such as listening to music, might not significantly impact pedestrian perception \cite{tian2022impacts}.

\paragraph{Decision} At uncontrolled crossings without signal lights, pedestrians often interact with yielding or non-yielding vehicles \cite{lee2022learning,lobjois2007age,pawar2022modelling}. In non-yielding scenarios, pedestrians usually make crossing decisions by evaluating gaps between approaching vehicles, known as gap acceptance behaviour (GA) \cite{theofilatos2021cross}. This concept led to the development of critical gap models, including the models by Raff \cite{raff1950volume}, HCM2010 \cite{manual2010hcm2010}, and Rasouli \cite{rasouli2022intend}. Alternatively, binary logit models treat crossing decisions as binary variables, utilising machine learning algorithms like Artificial Neural Networks (ANN), Support Vector Machines (SVM), and Logistic Regression (LR) \cite{himanen1988application}. For example,  Kadali et al. \cite{kadali2014evaluation} used ANN to predict crossing decisions based on various independent variables  (Figure \ref{figlabelkai2}c), while Sun et al. \cite{sun2003modeling} employed LR with variables such as pedestrian age, gender, group size, and vehicle type. 

In scenarios involving yielding vehicles, crossing decisions tend to follow a bimodal pattern referred to as bimodal crossing behaviour (BC) \cite{lee2022learning,pekkanen2022variable,Tian2023deceleration}. Pedestrians prefer to cross when traffic gaps are sufficiently large or when vehicles are about to stop \cite{pekkanen2022variable}. However, making decisions in such scenarios can be challenging due to the contrasting relationships between decision cues and collision risk, with collision risk negatively correlated to traffic gap and positively correlated to vehicle speed \cite{Tian2023deceleration}. Zhu et al. \cite{zhu2022defensive} clustered crossing decisions into three groups: crossing, dilemma condition, and waiting, based on vehicle speed and distance (Figure \ref{figlabelkai2}d). Moreover, Tian et al. \cite{tian2023psychological} assumed pedestrians applied different decision-making strategies according to BC behaviour and modelled crossing decisions as responses to different visual cues.

\label{crossref1}

While the mentioned approaches model crossing decisions based on observed behaviour patterns, other models delve into the psychological mechanisms that underlie these decisions. Specifically, Tian et al. modelled pedestrian GA behaviour based on pedestrian visual cues \cite{tian2022explaining} and extended it to yielding scenarios with a more complex visual perception mechanism \cite{tian2023psychological}. Wang et al. utilised a Reinforcement Learning (RL) model to capture pedestrian crossing behaviour based on limited perception mechanisms \cite{wang2023modeling}. Furthermore, a class of models, namely Evidence Accumulation (EA) models, such as the drift-diffusion model, proposed that crossing decisions result from the accumulation of visual evidence and noise, with the decision determined once a certain threshold was reached \cite{pekkanen2022variable,giles2019zebra,baker2019emergent,markkula2018models}. \cite{markkula2023explaining} integrated large-scale psychological theories to explain pedestrian crossing decisions in detail (Figure \ref{figlabelkai2}e). Additionally, game theory has also been applied to model crossing decisions when pedestrians negotiate the right of way with vehicles. Conventional game theory \cite{wu2019game}, Sequential Chicken (SC) game \cite{camara2022continuous}, and Dual Accumulator (DA) game \cite{kalantari2023modelling} were utilised to characterise the dynamic crossing decisions. 

Environmental variability and pedestrian heterogeneity further complicate crossing decision modelling. For example, crossing multiple lanes often involves pedestrians waiting at lane lines and accepting traffic gaps successively, known as rolling gap behaviour  \cite{zhang2018pedestrian,brewer2006exploration}. Pedestrians waiting at lane lines might be more likely to accept smaller traffic gaps than those waiting at curbs \cite{zhang2018pedestrian,zhao2019gap}. Another complex scenario is crossing a two-way road, which is physically and cognitively demanding. Pedestrians need to consider vehicles on both sides \cite{dommes2019street}. Similarly, crossing at intersections with dense continuous traffic is also challenging, as pedestrians need to anticipate crossing gaps upstream of traffic and make trade-offs between safety and time efficiency \cite{sucha2017pedestrian}. Typically, it was assumed that as waiting time increases, pedestrians tended to accept riskier crossing opportunities \cite{zhao2019gap,rasouli2022intend}. However, the latest evidence suggested that pedestrians who tended to wait were more cautious and less likely to accept risky gaps \cite{lobjois2013effects,tian2022impacts,yannis2013pedestrian}. Regarding pedestrian heterogeneity, ANN and LR models were applied to characterise age impact on crossing decisions by \cite{zhang2020pedestrian,kadali2014evaluation}. Distractions, such as cellphone usage, could also influence pedestrian crossing decisions \cite{liu2021modeling,thompson2013impact}. \cite{kadali2014evaluation} applied ANNs to model the impact of cellphone usage on crossing decisions. Furthermore, pedestrians often cross the road in a group, exhibiting herd behaviour. This behaviour was described by the tendency of group members to maintain a certain distance from the group centre \cite{moussaid2010walking}. \cite{tump2020wise} used an EA model to characterize information cascades in group decision-making, taking into account the influence of previous agents' decisions.

\begin{table}\normalsize
\centering
\caption{Pedestrian models and theories} 
\scalebox{1.0}{
\begin{tabular}{c c c c c}

\makecell[c]{\textbf{Research}} & \makecell[c]{\textbf{Scenario}} & \makecell[c]{\textbf{Cognitive process}} & 
\makecell[c]{\textbf{Models}} &\makecell[c]{\textbf{Theories}}  \\ \hline\hline

\makecell[c]{\cite{markkula2018models}} & \makecell[c]{1,3,4} & \makecell[c]{Perception} & \makecell[c]{$\tau$, $\dot{\tau}$} & \makecell[c]{Visual perception} \\ \hline

\makecell[c]{\cite{giles2019zebra}} & \makecell[c]{1,3,4} & \makecell[c]{Perception} & \makecell[c]{$\tau$, $\dot{\tau}$} & \makecell[c]{Visual perception} \\ \hline

\makecell[c]{\cite{tian2022explaining,tian2023deconstructing}} & \makecell[c]{1,3} & \makecell[c]{Perception} & \makecell[c]{$\dot{\theta}$} & \makecell[c]{Visual perception}  \\ \hline

\makecell[c]{\cite{domeyer2022driver}} & \makecell[c]{1,4} & \makecell[c]{Perception} & \makecell[c]{$\tau$,\\bearing angle} & \makecell[c]{Visual perception}  \\ \hline

\makecell[c]{\cite{pekkanen2022variable}} & \makecell[c]{1,3,4} & \makecell[c]{Perception} & \makecell[c]{Generalised TTC} & \makecell[c]{Visual perception} \\ \hline

\makecell[c]{\cite{predhumeau2022agent}} & \makecell[c]{1,4} & \makecell[c]{Perception} & \makecell[c]{TTC,\\ bearing angle} & \makecell[c]{Visual perception}  \\ \hline

\makecell[c]{\cite{wang2023modeling}} & \makecell[c]{1,3} & \makecell[c]{Perception} & \makecell[c]{Perceived distance} & \makecell[c]{Visual perception}\\\hline\hline

\makecell[c]{\cite{raff1950volume}} & \makecell[c]{1,3} & \makecell[c]{Decision} & \makecell[c]{Critical gap} & \makecell[c]{GA behaviour}  \\\hline

\makecell[c]{\cite{manual2010hcm2010} \\ \cite{pawar2016critical}} & \makecell[c]{1,3} & \makecell[c]{Decision} & \makecell[c]{Critical gap} & \makecell[c]{GA behaviour}   \\\hline

\makecell[c]{\cite{tian2023deconstructing}\\ \cite{zhao2019gap}} & \makecell[c]{1,3,5} & \makecell[c]{Decision} & \makecell[c]{LR} & \makecell[c]{GA behaviour}   \\\hline

\makecell[c]{\cite{kadali2014evaluation}} & \makecell[c]{1,3} & \makecell[c]{Decision} & \makecell[c]{ANN} & \makecell[c]{GA behaviour\\Machine learning}  \\\hline

\makecell[c]{\cite{rasouli2022intend}} & \makecell[c]{1,2,3,\\4,5} & \makecell[c]{Decision} & \makecell[c]{Critical gap} & \makecell[c]{GA behaviour} \\\hline

\makecell[c]{\cite{wang2023modeling}} & \makecell[c]{1,3} & \makecell[c]{Decision} & \makecell[c]{RL,\\Bayesian filter} & \makecell[c]{GA behaviour,\\learning-based}  \\\hline

\makecell[c]{\cite{zhu2022defensive}} & \makecell[c]{2,3,4} & \makecell[c]{Decision} & \makecell[c]{Speed-distance} & \makecell[c]{BC behaviour}  \\\hline

\makecell[c]{\cite{tian2023psychological}} & \makecell[c]{1,3,4} & \makecell[c]{Decision} & \makecell[c]{Hybrid perception,\\LR} & \makecell[c]{Visual perception,\\ BC behaviour}  \\\hline

\makecell[c]{\cite{markkula2018models},\\\cite{pekkanen2022variable}\\\cite{giles2019zebra}} & \makecell[c]{1,3,4} & \makecell[c]{Decision} & \makecell[c]{EA} & \makecell[c]{Drift diffusion}  \\\hline

\makecell[c]{\cite{markkula2023explaining}} & \makecell[c]{1,3,4} & \makecell[c]{Decision} & \makecell[c]{EA,\\Bayesian filter} & \makecell[c]{\scriptsize Drift diffusion, \\\scriptsize Game theory, \\\scriptsize Theory of Mind,\\\scriptsize Noisy visual perception} \\\hline

\makecell[c]{\cite{kalantari2023modelling}} & \makecell[c]{1,3,4} & \makecell[c]{Decision} & \makecell[c]{DA game} & \makecell[c]{Game theory}\\\hline

\makecell[c]{\cite{camara2022continuous}} & \makecell[c]{1,4} & \makecell[c]{Decision} & \makecell[c]{SC game} & \makecell[c]{Game theory} \\\hline

\makecell[c]{\cite{predhumeau2022agent}} & \makecell[c]{1,4} & \makecell[c]{Decision} & \makecell[c]{Critical gap} & \makecell[c]{GA behaviour,\\ visual perception}  \\\hline \hline

\makecell[c]{\cite{markkula2018models,markkula2023explaining},\\\cite{pekkanen2022variable}\\\cite{giles2019zebra}} & \makecell[c]{1,3,4} & \makecell[c]{Initiation} & \makecell[c]{EA} & \makecell[c]{Drift diffusion}  \\\hline

\makecell[c]{\cite{tian2023deconstructing}\\ \cite{tian2023psychological}} & \makecell[c]{1,3,4} & \makecell[c]{Initiation} & \makecell[c]{SW distribution} & \makecell[c]{Response time} \\ \hline 

\makecell[c]{\cite{wang2023modeling}} & \makecell[c]{1,3} & \makecell[c]{Initiation} & \makecell[c]{RL} & \makecell[c]{Learning-based}\\ \hline \hline

\makecell[c]{\cite{predhumeau2022agent}\\\cite{zeng2014application}\\\cite{yang2020social}} & \makecell[c]{1,3,4} & \makecell[c]{Motion} & \makecell[c]{SF} & \makecell[c]{Walking behaviour} \\ \hline

\makecell[c]{\cite{ma2016artificial}} & \makecell[c]{2} & \makecell[c]{Motion} & \makecell[c]{ANN} & \makecell[c]{Learning-based}  \\ \hline

\makecell[c]{\cite{layegh2020modeling} \\ \cite{lu2016cellular}} & \makecell[c]{1,2,3,4} & \makecell[c]{Motion} & \makecell[c]{CA} & \makecell[c]{Walking behaviour}  \\ \hline

\makecell[c]{\cite{nasernejad2022multiagent}} & \makecell[c]{2,3,4} & \makecell[c]{Motion} & \makecell[c]{Adversarial IRL} & \makecell[c]{Learning-based }  \\ \hline

\makecell[c]{\cite{kalatian2022context}} & \makecell[c]{2,4} & \makecell[c]{Motion} & \makecell[c]{LSTM} & \makecell[c]{Learning-based}  \\ \hline 

\multicolumn{5}{l}{\small 1. Uncontrolled crossings. 2. Controlled crossings. 3. With non-yielding vehicles. }\\ 

\multicolumn{5}{l}{\small 4. With yielding vehicles. 5. With traffic flow.}\\ 
 
\end{tabular} \label{tablekai1}
}
\end{table}

\begin{figure*}[t]
      \flushleft
     \includegraphics[width=1.0\linewidth]{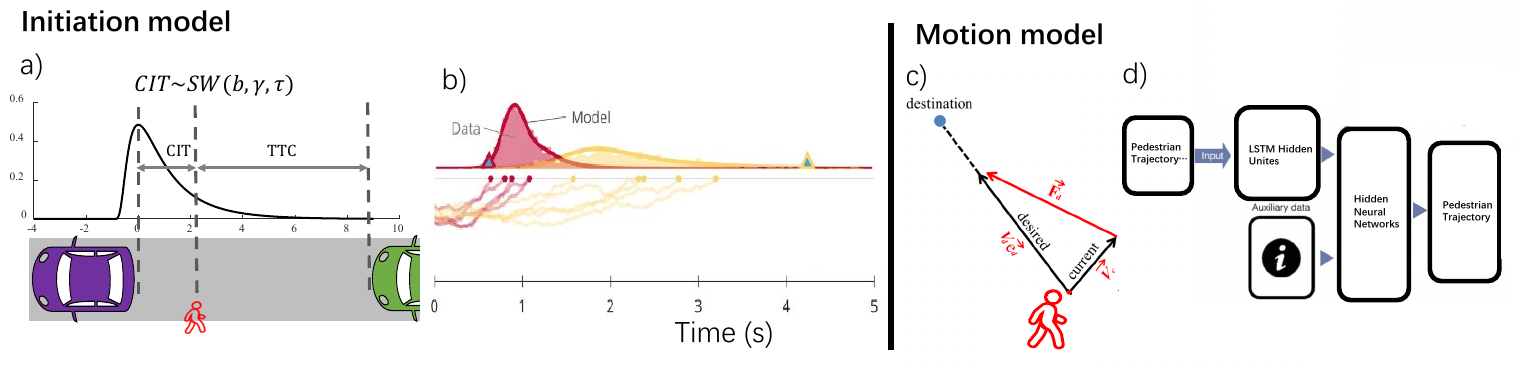}
      \caption{Initiation and motion models for pedestrians. a) Response time model \cite{tian2023deconstructing}. b) Evidence accumulation model \cite{markkula2018models}. c) Social force model \cite{zhang2020pedestrian}. d) LSTM-ANN \cite{kalatian2022context}. }
      \label{figlabelkai3}
\end{figure*}

\paragraph{Initiation and motion} crossing initiation time (CIT), represents the time it takes for pedestrians to begin crossing the road, reflecting the dynamic nature of their decisions \cite{domeyer2019proxemics}. Generally, CIT is the duration between the moment when the crossing opportunity is available and when the pedestrian begins to move \cite{tian2022explaining,lobjois2007age}. Drift-diffusion theory posited that CIT is influenced by the accumulation of noisy evidence in the cognitive system, reflecting the efficiency of pedestrian cognitive and locomotor systems  \cite{markkula2021accumulation}. Various factors could affect CIT, including vehicle kinematics, age, gender, and distractions. Pedestrians tended to initiate crossing more slowly when faced with higher vehicle speeds \cite{tian2022explaining}. Furthermore, female pedestrians tended to initiate crossings more quickly than males \cite{lobjois2009effects}, and the elderly tend to initiate sooner than young pedestrians \cite{lobjois2013effects}. The impact of distractions on the CIT depends on their components \cite{tian2022impacts}.

In scenarios where pedestrians face non-yielding vehicles, the risk of collision increases as the distance between the vehicle and the pedestrian decreases. Therefore, pedestrians typically make rapid decisions by assessing "snapshots" of approaching vehicles \cite{lee2022learning,lobjois2007age}. The distribution of CIT in these scenarios is often concentrated, and right-skewed  \cite{giles2019zebra}. Response time models, such as the Ex-Gaussian and Shifted Wald (SW) distributions, were used to model CITs in these situations  \cite{anders2016shifted}. For instance,  \cite{tian2022decision} modelled CITs as variables following SW distribution (Figure \ref{figlabelkai3}a). 

In vehicle-yielding scenarios, as discussed in Section \ref{crossref1}, CITs exhibit a bimodal distribution \cite{pekkanen2022variable}. For the early group of CITs, the distribution is similar to that in non-yielding scenarios, as pedestrians employ similar decision-making strategies\cite{Tian2023deceleration}. However, for the late group, the distribution is complex and cannot be described by standard response time distributions \cite{pekkanen2022variable}. EA models with time-varying evidence have been proposed to address this complexity, allowing for the generation of CIT distributions with intricate shapes \cite{pekkanen2022variable,giles2019zebra} (Figure \ref{figlabelkai3}b). Moreover, \cite{tian2023psychological} modelled CITs in vehicle-yielding scenarios using the joint distribution of response time models. Additionally, \cite{wang2023modeling} applied an RL model to learn the crossing initiation patterns of pedestrians.

After pedestrians initiate their crossings, they need to traverse the road. Walking is a key part of crossing behaviour and is influenced by many factors, such as the presence of approaching vehicles, infrastructures, pedestrian age, and distractions. Pedestrians adjusted their walking trajectories to avoid vehicles \cite{zeng2014application}. In multi-lane crossings, they tended to move to and wait at lane lines, accepting traffic gaps in each lane sequentially \cite{zhang2018pedestrian}. Pedestrian walking speeds at crossings were typically faster than normal walking speeds in other scenarios  \cite{montufar2007pedestrians}. While gender has no significant effect on walking speeds, teenagers and the elderly walk slower than young and middle-aged adults  \cite{montufar2007pedestrians,forde2021pedestrian}. Distractions, such as cellphone use, can reduce pedestrian walking speeds \cite{tian2022impacts}.

 Walking behaviour can be simulated using microscopic pedestrian motion models, including Cellular Automata (CA) models, Social Force (SF) models, and learning-based approaches. CA models are discrete in space, time, and state, making them ideal for simulating complex dynamic systems such as pedestrian-vehicle interactions \cite{lu2016cellular,layegh2020modeling}. The SF models, based on Newton's second law, were utilised to simulate pedestrian-vehicle interactions and large-scale pedestrian flows  \cite{helbing1995social,zeng2014application,moussaid2010walking} (Figure \ref{figlabelkai3}c). \cite{yang2020social} used an SF model to simulate the crossing behaviour of pedestrian crowds in complex interaction scenarios involving low-speed vehicles.

In contrast to the above \textit{white box} models, there are \textit{black box} models based on learning-based approaches, which learn pedestrian walking behaviour from naturalistic datasets or in pre-defined environments. For example, \cite{ma2016artificial} employed ANNs to learn pedestrian walking behaviour by incorporating the relative spatial and motion relationships between pedestrians and other objects extracted from videos. \cite{song2018data} used the outputs of an SF model as inputs to ANNs to simulate multiple pedestrian walking behaviours. \cite{kalatian2022context} proposed a Long Short-Term Memory Network (LSTM) pedestrian trajectory prediction model (Figure \ref{figlabelkai3}d). Additionally, RL and IRL models were also applied to model pedestrian walking behaviour. \cite{martinez2014marl} applied an RL model to learn multiple pedestrians' walking behaviour in an SF environment. \cite{nasernejad2022multiagent} developed an IRL model to learn pedestrian walking behaviour from video datasets.

\begin{table}[t]\normalsize
\centering
\caption{Applications of pedestrian theories and models in AV contexts} 
\scalebox{0.75}{
\begin{tabular}{lllll}
\makecell[c]{\textbf{Research}} & \makecell[c]{\textbf{Purpose}} & \makecell[c]{\textbf{Applied theory}} & \makecell[c]{\textbf{Applied model}}  \\ \hline \hline

\makecell[c]{\cite{li2021attentional}} & \makecell[c]{Pedestrian trajectory \\prediction } & \makecell[c]{Learning-based} & \makecell[c]{GCN} & \\ \hline

\makecell[c]{\cite{volz2016data}} & \makecell[c]{Pedestrian trajectory \\prediction } & \makecell[c]{Learning-based} & \makecell[c]{SVM, LSTM, \\ Dense NN} \\ \hline

\makecell[c]{\cite{kalatian2022context}} & \makecell[c]{Pedestrian trajectory \\prediction } & \makecell[c]{Learning-based} & \makecell[c]{LSTM} \\ \hline

\makecell[c]{\cite{zhang2020pedestrian}} & \makecell[c]{Pedestrian trajectory \\prediction } & \makecell[c]{GA behaviour,\\walking behaviour} & \makecell[c]{LR,SF} \\ \hline

\makecell[c]{\cite{predhumeau2022agent}} & \makecell[c]{Pedestrian behaviour \\modelling } & \makecell[c]{Visual perception,\\ GA behaviour,\\ walking behaviour} & \makecell[c]{Critical gap\\ SF \\bearing angle} \\ \hline

\makecell[c]{\cite{crosato2022interaction}} & \makecell[c]{Pedestrian behaviour\\modelling } & \makecell[c]{Crossing motivation,\\ walking behaviour} & \makecell[c]{LR,SF} \\ \hline

\makecell[c]{\cite{wang2023differentiated}} & \makecell[c]{Pedestrian behaviour \\modelling } & \makecell[c]{GA behaviour,\\ walking behaviour} & \makecell[c]{Critical gap, CA} \\ \hline

\makecell[c]{\cite{zhu2022defensive}} & \makecell[c]{Pedestrian behaviour\\ modelling } & \makecell[c]{BC behaviour} & \makecell[c]{Speed-distance} \\ \hline

\makecell[c]{\cite{camara2022continuous}} & \makecell[c]{Pedestrian behaviour \\modelling } & \makecell[c]{Game theory} & \makecell[c]{SC game} \\ \hline

\makecell[c]{\cite{domeyer2022driver}} & \makecell[c]{Pedestrian behaviour\\ modelling } & \makecell[c]{Visual perception, \\GA behaviour} & \makecell[c]{Critical gap \\ $\tau$ \\ bearing angle} \\ \hline

\makecell[c]{\cite{chen2023interaction}} & \makecell[c]{Pedestrian behaviour\\ modelling } & \makecell[c]{GA behaviour,\\ Game theory, \\ walking behaviour} & \makecell[c]{LR, \\ critical gap,\\ Stackelberg game,\\SF}  \\ \hline

\end{tabular} \label{tablekai2}
}
\end{table}

\subsubsection{AV-involved applications}
In recent years, there has been a growing interest in studying the interactions between Autonomous Vehicles (AVs) and pedestrians. This interest has led to a multitude of studies that apply pedestrian crossing behaviour theories and models to enhance or assess the performance of AVs in these interactions (Table \ref{tablekai2}). 

One prevalent approach is the use of learning-based methods, which learn pedestrian intention and trajectory from real-world datasets to aid AVs' decision-making. For instance, \cite{li2021attentional} proposed a Graph Convolutional Neural Network-based pedestrian trajectory prediction model, which considered past pedestrian trajectories to predict both deterministic and probabilistic future trajectories for a range of AV use cases. Other similar models aimed to improve prediction accuracy by considering the social context of interactions. For example, \cite{kalatian2022context} proposed an LSTM pedestrian trajectory prediction model, which considered past trajectories, pedestrian head orientations, and distance to the approaching vehicle as inputs \cite{rasouli2019autonomous}. In addition, there are studies aiming to anticipate pedestrian crossing intentions.  \cite{volz2016data} applied SVM, LSTM, and ANN to predict pedestrian crossing intentions separately.

Learning-based approaches have proven effective in predicting pedestrian trajectories and intentions. However, these models demand substantial data for robust performance, limiting their scalability when dealing with interaction cases lacking sufficient data. Moreover, the \textit{black box} nature of these models can make it challenging to interpret the generated trajectories and intents, which poses a challenge for AV decision-making modelling \cite{predhumeau2022agent}. To address these issues, expert models have been developed. For example, the SF model has been modified to predict pedestrian trajectories for AVs by incorporating more interaction details, such as TTC and the interaction angle between vehicles and pedestrians \cite{zhang2020pedestrian,predhumeau2022agent}. Moreover, SF and CA models have also been embedded in AV decision modules to represent pedestrian crossing behaviour and guide AV decisions in interactions with pedestrians \cite{crosato2022interaction,wang2023differentiated,chen2023interaction}.

Furthermore, crossing decision models have also been applied in AV research. For example, \cite{wang2023differentiated} employed crossing critical gap models to characterise pedestrian crossing decisions in their AV decision module. \cite{zhu2022defensive} applied their speed-distance model to design defensive and competitive interaction behaviour for AVs. \cite{zhang2020pedestrian,crosato2022interaction} used LR models as pedestrian crossing decision models in their proposed AVs decision-making modules. To enhance the dynamic and interactive nature of crossing decisions, game theoretical models were used to model crossing decisions when negotiating the right of way with AVs \cite{camara2022continuous,chen2023interaction}. 

Researchers also attempted to use pedestrian perception theories or models to design AV decision-making strategies. For instance, \cite{domeyer2022driver} simulated AV-pedestrian coupling behaviour using visual cues, $\tau$ and bearing angle, based on control theory. \cite{predhumeau2022agent} modelled the right of way of AVs and pedestrians using bearing angle.

\section{Interaction Modelling}
\label{sec:Interaction Modelling}
\justify

\begin{figure*}[t]
\centering
    \includegraphics[width=1.0\linewidth]{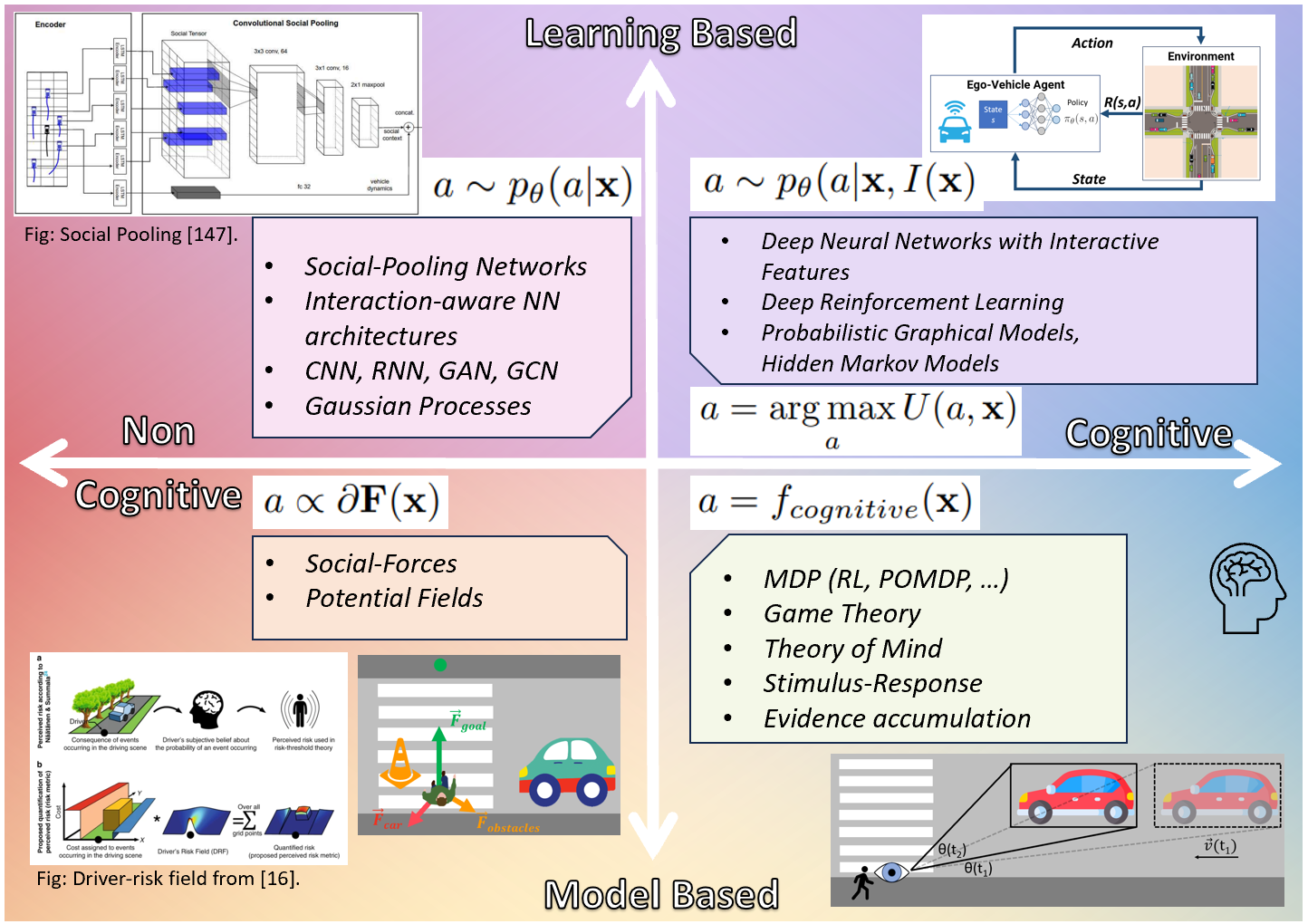}
    \caption{A map of state-of-the-art techniques in interaction-aware autonomous driving. }
    \label{fig:models}
    \vspace{-.0 cm}
\end{figure*}

Interaction modelling techniques are relevant to a huge variety of autonomous driving tasks, ranging from traffic forecasting to AV planning and decision-making. Understanding and modelling social interactions in autonomous driving is essential for predicting scene dynamics and ensuring safe AV behaviour. Accurate predictions enhance safety, while misunderstood AV behaviour can lead to accidents. Additionally, comprehending the social impact of AV actions can influence surrounding traffic, like encouraging pedestrian crossings through early stopping.
As interaction modelling techniques can be utilised across different task domains, the focus will be on dividing existing interaction modelling techniques regardless of the specific driving task they have been designed for. 

An introduction to the classification of interaction modeling techniques is now presented for discussion.
The first distinction can be made between learning-based methods and model-based methods. There has been extensive research in autonomous driving that makes use of machine-learning and deep-learning based techniques \cite{bachute2021autonomous}.
In a learning-based approach, a model is learned from an extensive dataset. This set of methods does not require any prior knowledge of the system. Data-driven methods are trained on a dataset of examples, and then they are used to make predictions or decisions. 
On the opposite side of the spectrum, model-based methods start with a theoretical understanding of the system. This a priori knowledge is used to create a mathematical model of the system. Empirical data is then used to validate the model or adjust its parameters to minimize the discrepancy between the model predictions and the data.

A further distinction can be made between methods that explicitly utilise cognitive features of the human mind, which try to explain the rationale that explains human actions, and methods that only implicitly try to model interactions, trying to map environmental inputs to decisions/actions. 
Human behaviour studies introduced in Section \ref{sec:Human Behaviour Studies} can serve as a guideline for the development of explicit methods.
For instance, game theoretic methods (see Section \ref{sec:Game Theory}) take a more explicit approach by considering traffic participants as rational agents who actively consider each other's actions. On the other hand, as an example of non-cognitive approaches, social force methods offer a more empirical perspective, capturing the impact of one participant on another without explicitly detailing the reasoning that explains the agent's behaviour during the interaction. We propose to distinguish existing modelling approaches based on whether they explicitly or implicitly model the interactions.

Based on these two criteria, four major categories of interaction modelling are identified, and they are reported in Figure \ref{fig:models}: 

\subsection{Learning-based Implicit Methods} 
These types of methods rely on machine learning or deep learning techniques. The interactions are implicitly modelled, which means that the agent's behaviour cannot be explained by the model. The model only learns an input-output mapping from the data. Model learning can be facilitated by exploiting interactive model architectures \cite{alahi2016social, xu2018encoding, bi2019joint, amirian2019social}.  In general, deep learning methods that use interaction-specific neural network architectures fall into this category.

In this type of method, the aim is to learn a probabilistic generative model that predicts the agent's future actions $a$. The model is a probability distribution conditioned on the environment state $\mathbf{x}$, which includes the state of surrounding agents, and a set of learnable parameters $\mathbf{\theta}$. 
\begin{equation} \label{eq: learning based implicit}
    a \sim p_{\mathbf{\theta}}(a|\mathbf{x}) 
\end{equation}

\subsection{Learning-based Methods with Cognitive Features} 
This set of methods relies on explicitly handcrafted interactive features that are used as inputs for a learning-based system. This type of interactive feature can include TTC, relative distance \cite{chen2021prediction}, looming and reflecting some cognitive process behind human reasoning. 
For example, in \cite{yoon2021design}, an LSTM which utilises the inter-vehicle interactions has been developed to classify surrounding vehicles' lane change intentions. The interaction features are composed of risk matrices which account for worst-case TTC with vehicles in surrounding lanes and relative distance. Graph Convolutional Networks also fall into this category, as interaction features can be explicitly modelled in the adjacency matrix of the graph \cite{rainbow21semantics,li22multiclasssgcn}.  

In this type of method, the aim is to learn a probabilistic generative model that predicts the agent's future actions $a$, similarly to \ref{eq: learning based implicit}. In this case, the probability distribution can be conditioned on the environment state $\mathbf{x}$ and on explicitly handcrafted interactive features $I(\mathbf{x})$, which have the purpose of facilitating learning.
\begin{equation} \label{eq: learning based explicit}
    a \sim p_{\mathbf{\theta}}(a|\mathbf{x}, I(\mathbf{x}) )
\end{equation}

\subsection{Model-based Non-Cognitive Methods} The modelling is non-cognitive in the sense that the interactions do not actively reason on the cognitive process that is behind the agent's actions. Methods of this group include SF \cite{yang2020social} and potential fields. The interactions are described by potential functions (or SF), which contain a set of learnable parameters which can be fit from empirical data. 
Another set of methods includes driver risk fields, which are based on the hypothesis that the driver behaviour emerges from a risk-based field \cite{kolekar2020human, li2020dynamic}. 
The advantage of model-based implicit methods is that they can be easily interpreted and they can embed domain knowledge, such as traffic regulations and scene context. Some models define a potential field and define the agent's action as proportional to the gradient of such field:
\begin{equation} \label{eq: model based implicit}
    a \propto \partial \mathbf{F}(\mathbf{x})
\end{equation}
Otherwise, the forces can be modelled directly so that the gradient operation is not required $a \propto \mathbf{F}(\mathbf{x})$.

\subsection{Model-based Cognitive Methods}
Model-based cognitive methods describe the reasoning behind human decision-making. Two main sets of methods can be distinguished: utility maximisation models and cognitive models.

In utility maximisation methods, humans are modelled as optimizers that select their actions so as to maximise their future utility. 
\begin{equation} \label{eq: utility maximisation}
     a = \argmax_a U(a, \mathbf{x})
\end{equation}
These methods include game theory and Markov Decision Processes (MDPs).
In Game theoretic approaches, agents are modelled as players competing or cooperating with each other, thereby taking into account how they react to each other \cite{wang2021socially, zhao2021yield}. The framework of game theory offers a transparent and clear-cut solution for modelling the dynamic interactions among human drivers, allowing for an understandable explanation of the decision process. However, it still remains hard to satisfy computational tractability as this approach does not scale well with an increasing number of agents.
Another possible solution is to model human behaviour as an agent of an MDP, which provides an excellent framework to model decision-making in scenarios where results are influenced by both chance and the decisions made by a decision-maker. Solutions to MDPs can be found with learning-based methods, e.g. DRL algorithms or Monte Carlo Tree Search \cite{chaslot2010monte}, or with dynamic programming techniques \cite{sutton2018reinforcement}.

The second set of methods aims to capture behavioural motivations behind agents' actions with psychological cognitive processes. This set of methods can include: 
\begin{itemize}
    \item \textbf{Stimulus-response} models  \cite{tian2022explaining}, where driver or pedestrian actions are determined, for example, on visual stimuli in the retina; 
    \item \textbf{Evidence accumulation} \cite{pekkanen2022variable}, where decisions are described as a result of accumulated evidence;
    \item \textbf{Theory of mind}, which suggests that humans use their understanding of others' thoughts and behaviours to make decisions. By predicting others' actions and inferring their knowledge, humans can drive effectively and safely \cite{sun2018probabilistic, chandra2021using}. 
\end{itemize}
\begin{equation} \label{eq:cognitive model}
    a = f_{cognitive}(\mathbf{x})
\end{equation}
\newline
In the next sections, each of these classes of interaction modelling will be analyzed in greated detail. In particular, Cognitive and Non-Cognitive learning-based methods will be discussed in Section \ref{sec:Learning Based Methods}. Model-based cognitive methods have already been thoroughly discussed in Section \ref{sec:Human Behaviour Studies}, where Social Force and Potential Fields, Driver Risk Field models, Theory of Mind, Stimulus-Response, and Evidence Accumulation models were included. Section \ref{sec:Utility Based Methods} will include Utility-Based methods, which comprise MDPs (Section \ref{sec:MDP}) and Game Theory (Section \ref{sec:Game Theory}).

\section{Learning Based Methods} 
\label{sec:Learning Based Methods}
\begin{figure*}[t]
\centering
    \includegraphics[width=1.0\linewidth]{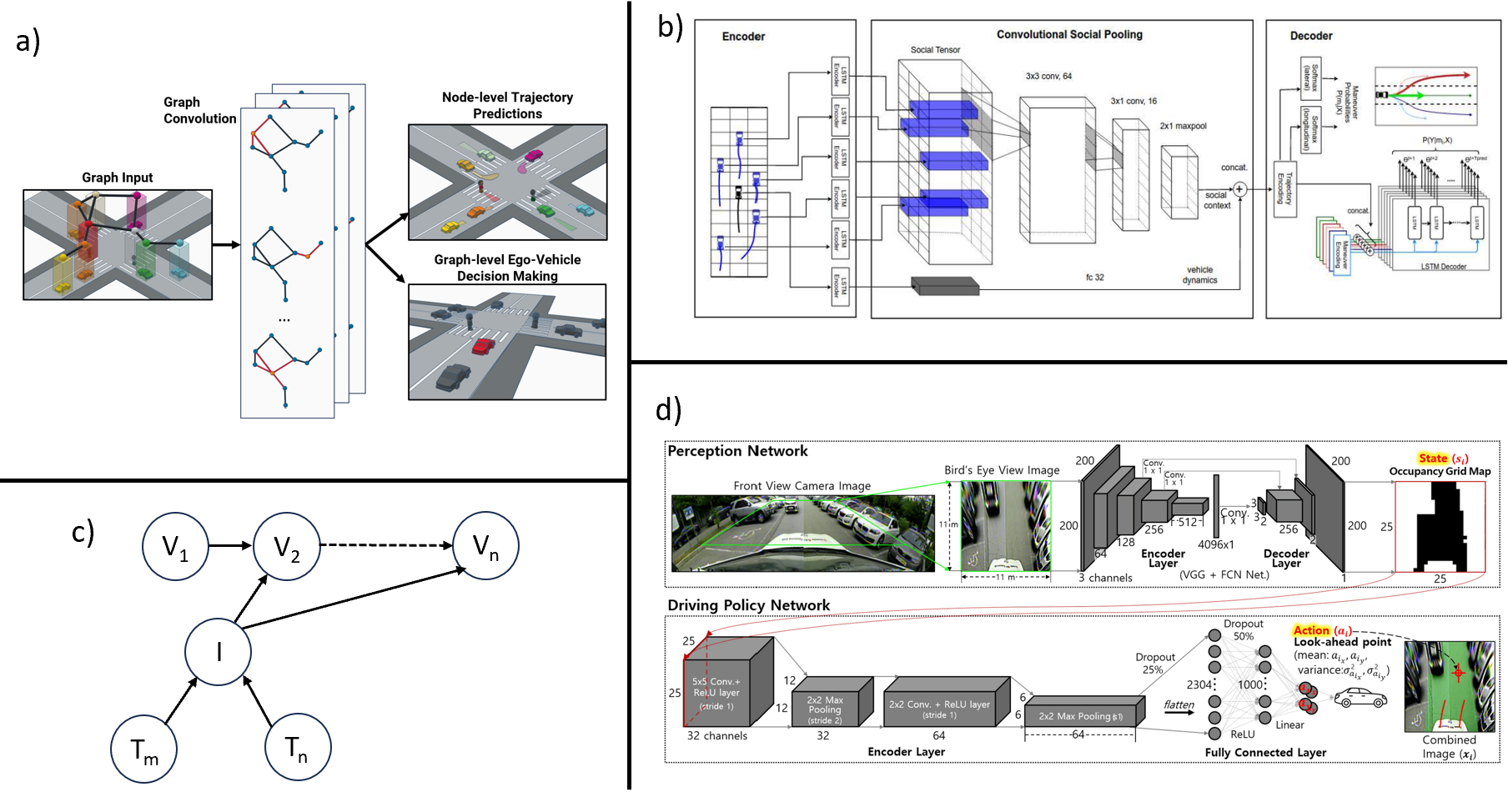}
    \caption{Overview figure of deep learning methods in interaction-aware tasks. a) GCNs can be used for both node-level predictions of surrounding agents behaviour as well as ego-vehicle motion generation (graph-level output), b) social-pooling operation in \cite{deo2018convolutional},  c) probabilistic graphical model in \cite{dong2017intention}, d) end-to-end imitation learning network in  \cite{ahn2022autonomous}.}
    \label{fig:learning-based}
    \vspace{0.0 cm}
\end{figure*}

\justify
Machine Learning (ML) methods are widely used in autonomous driving for a variety of tasks, including object detection \cite{feng2021review}, scene understanding \cite{guo2021survey}, path planning and control \cite{aradi2020survey}. By learning from large amounts of data, ML methods can learn to make decisions that are more accurate and efficient than those made by humans \cite{mnih2015human}. This Section will comprise both implicit and explicit learning-based methods identified in the previous Section and give a more detailed view of relevant papers. An overview of some learning-based methods is shown in Figure \ref{fig:learning-based}.

Thanks to recent improvements in neural networks learning representations, it is now possible to use \textit{end-to-end} driving approaches that take as input the raw sensor readings to output control commands, such as steering and throttle, to solve path-planning and control problems \cite{hu21multitask}. 
However, it is challenging to learn the entirety of the driving task from high-dimensional raw sensory data (e.g. LiDAR point clouds, camera images) as this involves learning perception and decision-making at the same time. In most of the works, the \textit{how-to-act} learning process assumes that a scene representation is available to the motion-planning and decision-making module. This actually requires splitting the end-to-end driving into two main blocks, one in which the AV learns \textit{how-to-see} and one in which it learns \textit{how-to-act}. 

There are two main approaches to end-to-end self-driving for planning and control tasks (\textit{how-to-act}):  
\begin{itemize}
    \item \textbf{Imitation Learning}: in which an agent learns to mimic the behaviour of an expert \cite{sun2018fast, cai2020dignet, bronstein2022hierarchical}. 
    \item \textbf{Deep Reinforcement Learning (DRL)}: in which an agent tries to learn how to act in a trial-and-error process that typically takes place in a simulated environment. DRL methods will be analysed in greater detail in Section \ref{sec:Utility Based Methods}.
\end{itemize}

Imitation learning is a machine learning paradigm in which an agent learns to perform tasks by imitating the behaviour of expert demonstrators, making it a valuable approach for training autonomous systems and robots. In \cite{cai2020dignet}, interactive features are learned by means of a Graph Attention Network (GAT). The input to this network consists of surrounding agents' kinematic information as well as a feature vector that encodes scene representation coming from a Bird's Eye View. The model is trained on synthetic data generated by an expert driver in the CARLA simulator. Imitation learning methods tend to work really well in scenarios that are similar to the training scenarios but typically fail when the scenarios diverge from the training distribution. Algorithms like Dataset Aggregation (DAgger) \cite{ross2011reduction} can improve the performances of imitation learning policies by augmenting the initial training dataset with human-labelled data for unseen situations. However, asking an expert to label new training samples can be expensive or unfeasible. 

In the context of scene understanding and motion prediction, deep neural networks have been extensively used.   \cite{alahi2016social} et al. proposed a social-pooling operation in their neural network architecture to account for surrounding neighbours in crowd motion prediction. Similarly, \cite{zhu2021star} made use of a star-topology network with max-pooling operation to account for interaction features in multi-agent forecasting. 
CIDNN \cite{xu2018encoding} uses LSTM to track the movement of each pedestrian in a crowd and assigns a weight to each pedestrian's motion feature based on their proximity to the target pedestrian for location prediction. The study in \cite{bi2019joint} created a dataset and proposed a framework called VP-LSTM to predict the trajectories of vehicles and pedestrians together in crowded mixed scenes by exploiting different LSTM architectures for heterogeneous agents. A Generative Adversarial Network (GAN) is applied in \cite{amirian2019social} to sample plausible predictions for any agent in the scene. 
The shared feature of these methods is the usage of Recurrent Neural Networks that capture spatio-temporal interaction features in conjunction with pooling operations. The pooling operation allows one to account for surrounding agents by mixing up the hidden states extracted by the LSTMs. During the social-pooling operation, the hidden states of surrounding agents become features that are used to predict the current agent motion. 
Diffusion models are another set of deep-learning techniques with increasing popularity in modelling spatial-temporal trajectories, which can be used for predicting both pedestrian and car trajectories \cite{chang23design}. 

Graph Convolutional Networks (GCNs) have been widely used in trajectory prediction tasks with interacting agents. 
In these methods, the road structure is represented as a graph, with each node representing a traffic participant. Each node can carry information such as the traffic participant's class (car, truck, pedestrian, etc.), its location, or speed. Explicit interaction can be modelled in the Adjacency Matrix of the graph, whereas the implicit part consists of the graph convolutional layers. GCNs are widely used in traffic forecasting \cite{lu2022vehicle, li2019grip, casas2020spagnn, mo2021graph}, and have also been recently used in motion planning \cite{chen2021graph, jin2022conquering, hart2020graph, hugle2020dynamic}, especially in combination with DRL.

Other machine learning techniques that can be used to model interactions include Gaussian Processes \cite{su2017forecast} and probabilistic graphical models, including Hidden Markov Models \cite{liu2020driving, jin2020gauss}. 

\section{Utility Based Methods}
\label{sec:Utility Based Methods}
\justify
Utility-based agents \cite{garrido2022review} employ utility functions to guide decision-making, assigning values to possible world states and selecting actions leading to the highest utility.  In contrast to goal-based agents, which evaluate states based on goal satisfaction, utility-based agents can handle multiple goals and factor in probability and action cost. Utility based methods encompass Markov Decision Processes (MDPs) and Game Theoretic Models which will be analysed in Sections \ref{sec:MDP} and \ref{sec:Game Theory}.

\subsection{Markov Decision Processes} \label{sec:MDP}
MDPs are a mathematical framework used to model decision-making problems where the outcomes are partly random and partly under the control of a decision-maker.
\begin{figure}[pbth]
\centering
    \includegraphics[width=.7\linewidth]{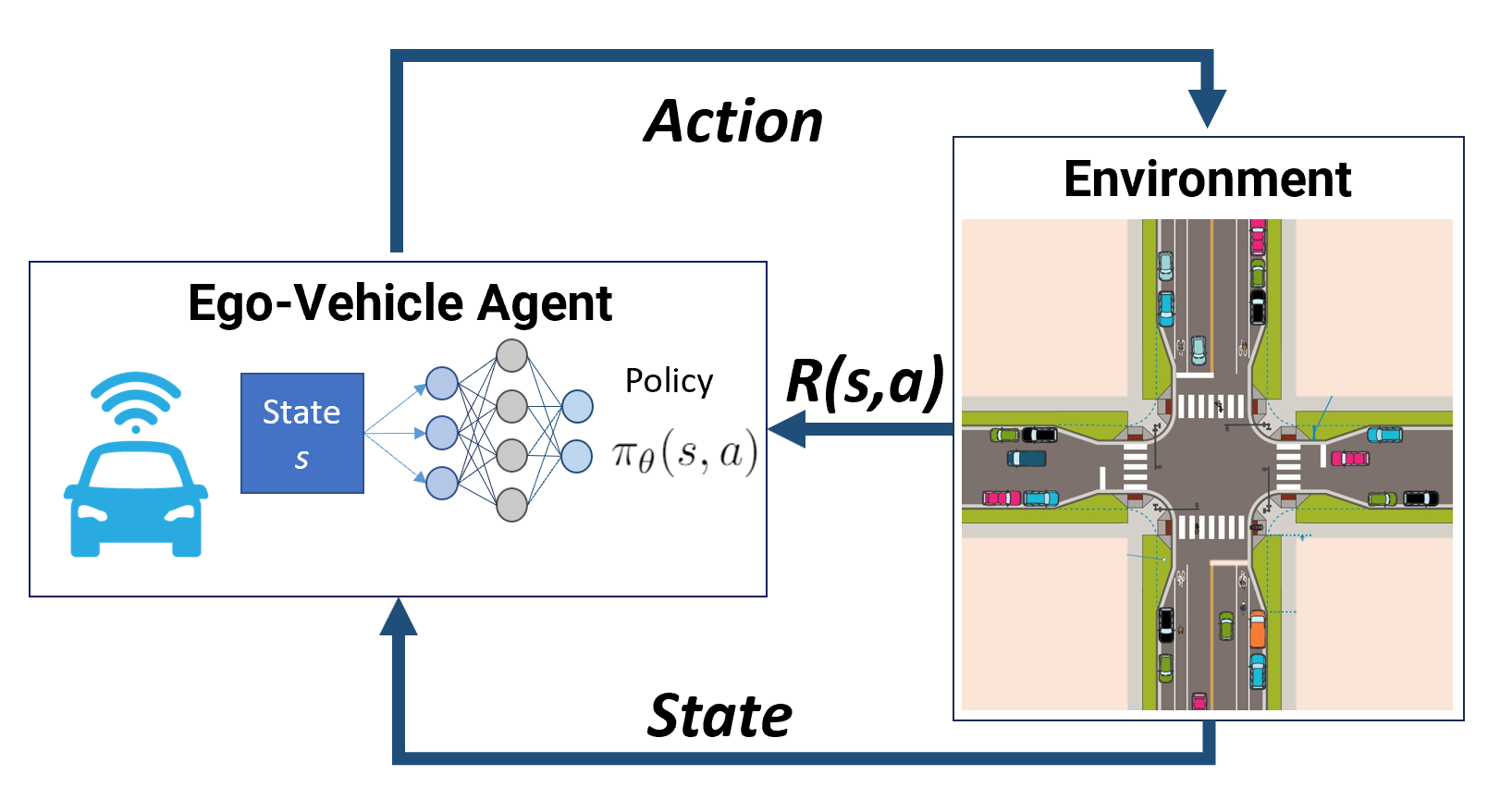}
    \caption{MDP framework. An agent takes an action that affects the environment state. The updated environment state is used to take the next action, and the cycle repeats. The reward function is used to define the objective of the MDP, which is to maximise the expected cumulative reward over time.}
    \label{fig:mdp}
    \vspace{-.0 cm}
\end{figure}
The modelling framework for MDPs is illustrated in Figure \ref{fig:mdp}. Two main methods exist to solve MDPs: dynamic programming and reinforcement-learning \cite{sutton2018reinforcement}. Typically the latter set of methods are more used in autonomous driving, as they are more suitable for high-dimensional state spaces.

\subsubsection{Reinforcement Learning} \label{sec:DRL}
Reinforcement Learning (RL) leverages Markov Decision Processes (MDPs) to model complex environments and comprises a set of algorithms to learn policies that maximise the expected reward \cite{sutton2018reinforcement}. 

Traditionally, Dynamic Programming is a reliable approach for this purpose, iteratively calculating the value of each state, commencing from terminal states and working backwards to the initial state. This method excels in scenarios with modest state spaces. However, it can be computationally burdensome when confronting RL challenges characterized by vast state spaces, such as the domain of Autonomous Driving. More commonly, RL augmented with deep neural networks (DRL) is used. DRL algorithms can be more sample-efficient and scalable than dynamic programming algorithms, but they can also be more complex and difficult to train. 
For a more detailed survey on DRL applications to autonomous driving, please refer to \cite{aradi2020survey}.

DRL solutions in autonomous driving will be classified based on the scenario used, the state space representation, the action space, and the algorithm used.

\begin{figure}[tpbh]
\centering
    \includegraphics[width=1.0\linewidth]{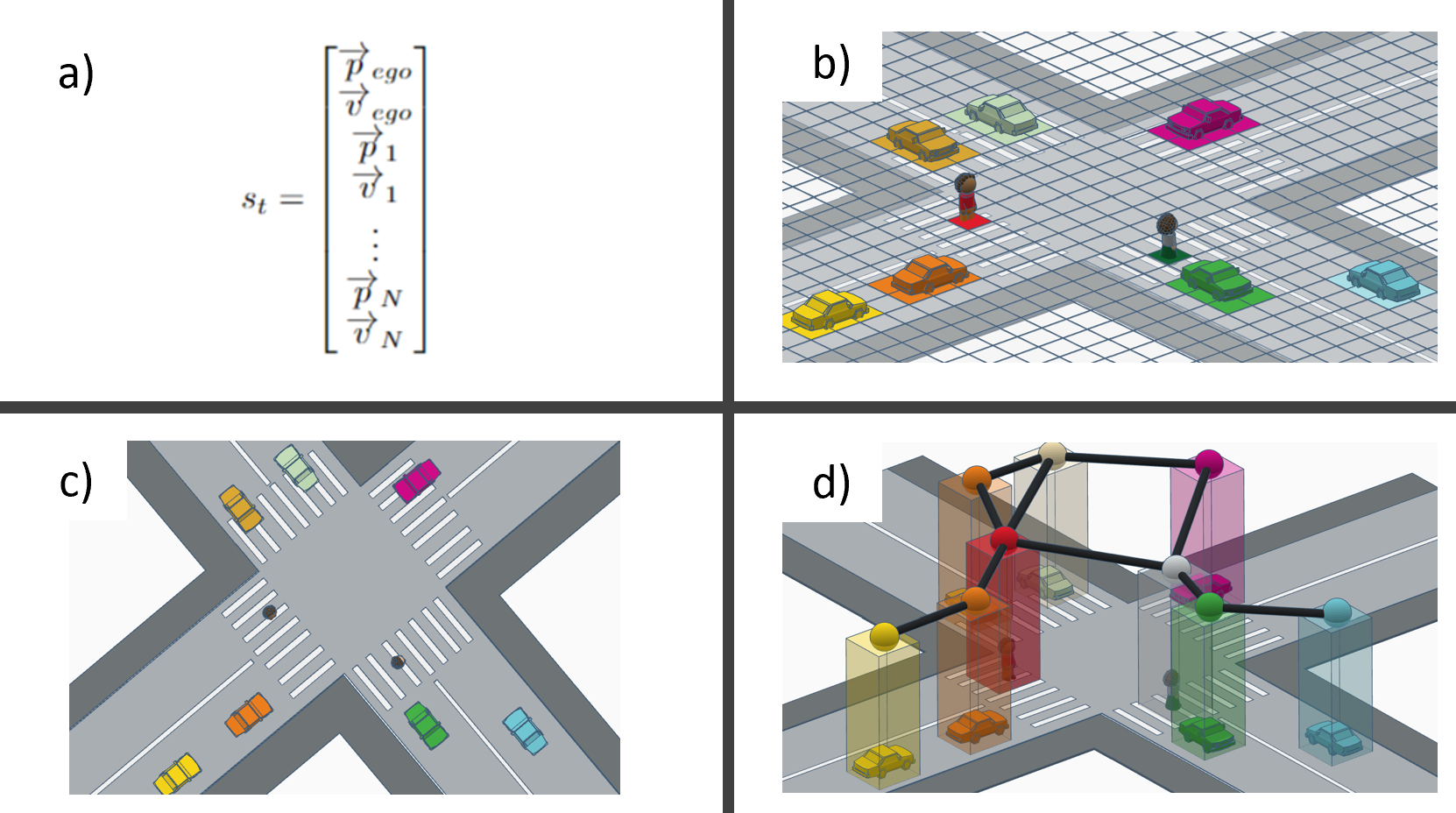}
    \caption{Illustration of state representations typically used in AD. a) vector representation, b) grid-based, c) Bird's Eye View, d) graph.}
    \label{fig:state_repr_rl}
    \vspace{-.0 cm}
\end{figure}

Typical state representations used in DRL include see Figure \ref{fig:state_repr_rl}:
\begin{itemize}
    \item \textit{Vector based representation}: in this type of representation, information regarding surrounding vehicles, such as position and velocity, is included in a vector of fixed length \cite{hoel2018automated};
    \item \textit{Bird's Eye View (BEV) Image}: a 2D image representation of the environment surrounding the ego-vehicle from a top-down perspective \cite{palanisamy2020multi};
    \item \textit{Occupancy grid representation}: similar to a BEV image, it is a 2D discrete representation of the environment that surrounds the ego-vehicle. It is a 2D or 3D grid of cells, each of which is assigned a probability of being occupied by an obstacle, as well as segmentation information regarding what type of entity is occupying the cell \cite{hu2019interaction, deshpande2019deep}.
    \item \textit{Graph representation}: it is a way of representing the state of the environment around an AV as a graph. The nodes in the graph represent objects in the environment, such as vehicles, pedestrians, and traffic lights. The edges in the graph represent the relationships between objects, such as proximity or potential for collision. Graph representations are compact and efficient and are a promising approach to representing the state of the environment \cite{hugle2020dynamic, hart2020graph}.
\end{itemize}
Vector-based representation offers a compact and efficient representation of objects at the expense of limiting traffic information to a subset of fixed dimensions of surrounding vehicles.
BEV images and occupancy grids offer a simple way to represent the environment with fixed and can be easily updated. However, they can be inaccurate in environments with high clutter or uncertainty. Graph representation can easily represent the relationships between agents in a compact way. On the other hand, it can be complex and computationally expensive to update the graph as the number of surrounding agents increases. \\

The action space can be continuous or discrete. Continuous actions usually include the ego vehicle's longitudinal acceleration and steering angle \cite{saxena2020driving}. Discrete actions usually depend on the specific task being solved. For example, in a lane change scenario, discrete actions include left-lane change, keeping the current lane, or right-lane change. The lower-level controller regulates the steering and acceleration of the vehicle to execute the manoeuvre \cite{deshpande2019deep, yuan2021deep}. 

Whilst most DRL papers focus on vehicle-only traffic scenes, the number of papers that deal with mixed traffic scenarios or with vehicle-pedestrian interaction is more limited. 
Some works exist in the mobile robots' crowd navigation. In \cite{everett2018motion}, DRL is used to navigate a robot in a crowd in a multi-agent setting. In \cite{chen2019crowd}, the model in \cite{everett2018motion} is improved by using attention-based neural networks and social pooling. 
An autonomous braking system was developed in \cite{chae2017autonomous} with a DQN agent. The authors implement a Trauma Memory, which is used to sample from collision scenarios in a way similar to Prioritized Experience Replay (PER) \cite{schaul2015prioritized}. 
In \cite{li2020deep}, a DQN agent is trained to avoid collisions with a crossing pedestrian and is further used to develop an ADAS system to aid drivers in pedestrian collision avoidance scenarios. Deshpande et al. \cite{deshpande2019deep, deshpande2020behavioral} used a grid-state representation with four layers.
In a similar scenario, the authors in \cite{crosato2021human} developed a SAC agent with continuous actions is used. By integrating SVO component in the reward function, the vehicle can be trained to have different social-compliant behaviours, from pro-social behaviours to more aggressive ones. \\

Deploying Deep Reinforcement Learning (DRL) in real-world scenarios poses a significant challenge and is an open research field.  Some studies, like \cite{seong2021learning}, implement DRL policies directly in real-world applications without additional fine-tuning, showcasing their effectiveness in scenarios like unsignalised intersections.
Transfer learning, a sub-field of deep learning, is currently being explored to transfer knowledge from the simulation to the real world. Two main techniques include domain adaptation and domain randomisation \cite{zhou2022domain}. With domain randomisation,  the approach aims to have a sufficiently large training dataset that encompasses the real world as a specific case \cite{tobin2017domain}. With domain adaptation \cite{shan2019pixel}, the aim is to learn from a source distribution a model that performs well on a target distribution. 

Another issue related to DRL is that the learning-based strategy has high training costs and makes it difficult to achieve semantic interpretation. Recently, some researchers have focused on interpretable learning algorithms and lifelong learning algorithms to solve the above shortcomings \cite{chen2021interpretable}.

\subsubsection{Multi Agent Reinforcement Learning}
\label{ssec:MARL}
When multiple RL agents are being deployed into the real world and interact with each other, the problem becomes Multi-Agent Reinforcement Learning (MARL). 
In order to deal with multi-agent systems, multiple approaches are possible. The first approach is to have a centralised controller that manages the entire fleet. By increasing the state dimension to include all vehicles and having a joint action vector, the problem can again become a single-agent problem \cite{klimke2022cooperative}. The drawback is the increased dimensionality of the state and action spaces, which can make learning more complex. Recently, graph-based representation has been employed to overcome the curse of dimensionality of the problem \cite{klimke2022cooperative}. 

Another approach, which takes inspiration from level-k game theory, is to have a single DRL learner but replace some of the surrounding agents with previous copies of itself \cite{yuan2021deep}. This technique is similar to self-play, which is used in competitive DRL scenarios \cite{baker2019emergent}. 
Finally, the last approach is to formulate the problem with a MARL approach, where multiple learners are in parallel. 
A multi-agent deep deterministic policy gradient (MADDPG) method is proposed in \cite{lowe2017multi}, which learns a separate centralized critic for each agent, allowing each agent to have different reward functions. See \cite{zhang2021multi} for an extensive review of MARL. Other applications of MARL in autonomous driving can be found in \cite{hu2019interaction, palanisamy2020multi,chen2021graph,chen2021midas}.

\subsubsection{Partially Observable Markov Decision Processes} \label{sec:POMDP}
Partially Observable Markov Decision Processes (POMDPs) are a generalisation of MDPs. If the process state $s$ cannot be directly observed by the decision-maker, the MDP is said to be Partially Observable. 
POMDPs are computationally expensive but provide a general framework that can model a variety of real-world decision-making processes. POMDP applications for Autonomous Driving are becoming increasingly popular thanks to hardware improvements. In \cite{luo2018porca}, POMDP has been used to navigate a mobile robot in a crowd. The robot keeps a belief over possible future goals for pedestrians. POMDPs have also been utilised for car decision-making in the presence of pedestrians \cite{hsu2020pomdp}. 
In a POMDP, the agents surrounding the ego-vehicle are modelled as part of the environment, and a belief vector is used to model their intentions. In \cite{chen2021midas}, the authors develop a multi-agent interaction-aware Decision-making policy with DRL.  The problem is modelled as a POMDP, and the interaction is modelled with an attention-based neural network mechanism.
POMDPs have also been used to solve decision-making problems under environmental occlusions at intersections \cite{hubmann2019pomdp}. Other applications of POMDP to interactive decision-making can be found in \cite{bouton2017belief, pusse2019hybrid}.
Traditional control methods usually deal with sensor uncertainty and planning in a sequential manner, where the sensor noises and uncertainties are dealt with by a state-estimator, and then a deterministic policy is used to determine the action based on the estimated state. POMDPs, on the other hand, do not make such separation, and the policy is determined based on a belief state.
Surrounding agents can be either explicitly modelled as decision-makers (MARL) or can be treated as part of the environment in which a single agent operates (RL or DRL) \cite{naveed2021trajectory, saxena2020driving}.


\begin{table}[t]\normalsize
\centering
\caption{DRL overview table} 
\scalebox{0.67}{
\begin{tabular}{llllllll}
\textbf{Research} & \textbf{Scenario} & \textbf{Observation} & \textbf{Action} & \textbf{Reward} & \textbf{DRL Alg} & \textbf{Network} & \textbf{Simulator} \\ \hline \hline

\makecell{\cite{li2020deep}} & \makecell{Pedestrian\\Collision Avoidance} & \makecell{Continuous\\($x_p$, $y_p$, $v$)} & \makecell{Discrete (break,\\ keep, change lane)} & \makecell{coll, smooth,\\succ} & DQN & MLP & PreScan \\ \hline

\makecell{\cite{crosato2021human}}& \makecell{Pedestrian\\Collision Avoidance} & \makecell{Continuous\\($x_p$, $y_p$, $v_p$, $v$)}  & Continuous ($a$) & \makecell{coll, speed,\\succ, SVO} & SAC & MLP & Custom \\ \hline

\makecell{\cite{hoel2018automated}} & Highway Navigation & \makecell{Continuous ($x$, $y$, $v$)\\ + 8 surr. veh.} & \makecell{Discrete \\(LLC, RLC, KEEP) or\\(LLC, RLC, 4*ACC)}& \makecell{coll, near coll,\\ lc, speed} & DQN & CNN & Custom \\ \hline

\makecell{\cite{chae2017autonomous}} & \makecell{Pedestrian\\Collision Avoidance} & \makecell{Continuous\\($x_p$, $y_p$, $v$)} & Discrete (ACC*4) & \makecell{coll,\\early break penalty} & DQN & MLP & PreScan \\ \hline 

\makecell{\cite{deshpande2020behavioral}}& \makecell{Pedestrian\\Collision Avoidance} & \makecell{Grid based\\representation} & \makecell{Discrete\\(ACC*4)}& \makecell{speed, coll,\\near coll} & DQN & \makecell{CNN\\+LSTM} & Custom \\ \hline

\makecell{\cite{yuan2021deep}} & Intersection & \makecell{Continuous\\($v$, path)\\+ 4*surr. veh.} & \makecell{Discrete\\(5*ACC)} & \makecell{coll, near coll, \\acc, succ}& \makecell{D3QN\\PER} & MLP & SUMO \\ \hline

\makecell{\cite{saxena2020driving}} & \makecell{Dense Traffic\\Lane Change} & \makecell{Grid}  & \makecell{Continuous\\(jerk, $\omega$)} & \makecell{coll, speed,\\jerk, time} & PPO & CNN & Custom \\ \hline
 
\makecell{\cite{hugle2020dynamic}} & Lane Change & \makecell{Graph\\($p_{rel}$, $v_{rel}$,\\lane index)} & \makecell{Discrete\\(LLC, RLC, KEEP)} & \makecell{speed, lc,\\coll} & DQN & GCN & SUMO \\ \hline

\makecell{\cite{hart2020graph}} & \makecell{Highway\\Lane Change} & \makecell{Graph\\($p_{rel}$, $v_{rel}$)} & \makecell{Continuous\\($\omega$, $a$)} & \makecell{coll, goal,\\vel, acc} & PPO & GCN & BARK \\ \hline

\makecell{\cite{li2021safe}} & \makecell{Intersection,\\Roundabout} & \makecell{k-nearest,\\($p_{rel}$, $v_{rel}$, $\theta$)} & Discrete & \makecell{distance,\\coll} & \makecell{H-CtRL\\DDQN} & MLP & Custom \\ \hline

\makecell{\cite{chen2021interpretable}} & Town Navigation & \makecell{Lidar,\\Camera} & \makecell{Continuous\\($a$, $\delta$)} & \makecell{coll, speed\\acc, time,\\speed limit,\\out lane} & Proposed & CNN & CARLA \\ \hline

\makecell{\cite{chen2019crowd}} & Crowd-navigation & \makecell{k-nearest\\($p_{real}$, $v_{rel}$))} & \makecell{Discrete 80\\(5*ACC, 16*angles)} & \makecell{coll, goal,\\proximity} & \makecell{Deep V-\\learning} & \makecell{Social-\\Attentive} & Custom \\ \hline

\makecell{\cite{everett2018motion}}  & Crowd-navigation & \makecell{k-nearest\\($p_{rel}$, $v_{rel}$, $d$, $r$} & \makecell{Discrete 11\\(angle-speed comb.)} & \makecell{coll, goal,\\proximity} & \makecell{GA3C-\\CADRL} & LSTM & Custom \\ \hline

\makecell{\cite{chen2021midas}} & \makecell{MARL\\Intersection,\\Roundabout} & all within range & - & \makecell{time, speed,\\coll, front-car\\distance} & Double DQN & Attention & Custom \\ \hline

\makecell{\cite{chen2021graph}} & \makecell{MARL\\Highway Navigation} & \makecell{($p_{rel}$, $v_{rel}$,\\ lane, intention)}& \makecell{Discrete 3\\(LLC, RLC, keep)} & \makecell{goal, speed,\\coll., lc} & \makecell{Graph Q\\(Proposed)} & GCN & SUMO\\ \hline
 
\makecell{\cite{hu2019interaction}} & \makecell{MARL\\Merge} & Cell grid & \makecell{Discrete\\(ACC*5)} & \makecell{goal, coll,\\flow} & Curriculum & MLP & Custom\\ \hline
 
\makecell{\cite{palanisamy2020multi}} & \makecell{MARL\\Connected AD} & RGB BEV Image & \makecell{Discrete 9\\Combinations of\\(Brake, Steer,\\Throttle)} & \makecell{goal, speed,\\coll, lc} & IMPALA & CNN & CARLA \\ \hline

\makecell{\cite{klimke2022cooperative}} & \makecell{MARL\\Connected AD} & \makecell{Graph\\($s$, $v$, $d$)} & \makecell{Continous\\($a$)} & \makecell{speed, action,\\idle, proximity,\\coll} & TD3 & GCN & \makecell{Highway-env\\Open Source} \\ \hline

 &  &  &  &  &  &  & \\ \hline \hline
\end{tabular}
}
\end{table}

\subsection{Game Theoretic Models} \label{sec:Game Theory}
Game theory is the study of mathematical models of strategical interactions between rational agents \cite{bacsar1998dynamic}. Game theory primarily applies to economics but has also emerged in autonomous driving. In particular, dynamic non-cooperative game theory is of particular importance for Autonomous Driving \cite{sadigh2018planning}. Game theory is \textit{dynamic} if it involves multiple decisions and the order of such decisions is important, and it is \textit{non-cooperative} if each person involved pursues their own interest, which is partly conflicting with other people's interest.  
The dynamic non-cooperative game theory comprises both discrete and continuous time games, and it provides a natural extension of optimal control to multi-agent settings \cite{bacsar1998dynamic}.

\begin{table}\normalsize
\centering
\caption{Game Theory Models for Decision Making in Various Scenarios} 
\scalebox{.9}{
\begin{tabular}{lllll}
\textbf{Research}   & \textbf{Scenario}    & \textbf{Game Theory Model}   & \textbf{Agents} & \textbf{Action}  \\ \hline \hline
\cite{wang2015game} & Lane Change & Nash Equilibrium    & Up to 4         & Both discrete (Lane \\
&&&&change) and continuous $a$ \\ \hline
\makecell{\cite{williams2018best}} & Autonomous Racing    & Nash Equilibrium & 2               & Continuous      \\ \hline
\makecell{\cite{fridovich2020efficient,fridovich2020iterative}} & Intersection & Nash Equilibrium  & 3  & Continuous ($\omega$ and $a$) \\ \hline
\makecell{\cite{spica2020real}}& Drone Racing     & Nash Equilibrium          & 2 & Continuous ($\omega$ and $v$)                       \\ \hline
\makecell{\cite{wang2021game}}& Autonomous Racing  & Nash Equilibrium          & 2 & Continuous ($a$ and $\delta$)      \\ \hline
\makecell{\cite{schwarting2021stochastic}}   & Autonomous Racing & Nash Equilibrium  & 2     & Continuous ($a$ and $\delta$)      \\ \hline
\makecell{\cite{schwarting2019social}}& Merging            & Nash Equilibrium, SVO   & 2       & Continuous ($a$ and $\delta$)     \\ \hline
\makecell{\cite{ding2018game}}& Highway, Lane Change & Nash Equilibrium, DRL      & 2      & Discrete ($a$ and lane change)  \\ \hline
\makecell{\cite{liniger2019noncooperative}}& Autonomous Racing    & Nash Eq. and Stackelberg Eq. & 2   & Discrete trajectories\\ \hline
\makecell{\cite{sadigh2018planning}} & Highway navigation, & Stackelberg Eq. &  2 &   Continuous ($a$ and $\delta$) \\ 
&Intersection&&&\\
\hline
\makecell{\cite{hang2020integrated}}& Lane Change &  Stackelberg Eq. & 3 & Both discrete and continuous \\ \hline
\makecell{\cite{yoo2020stackelberg}} & Lane Change &  Stackelberg Eq. & 2 & Discrete (Lane change) \\ \hline
\makecell{\cite{wang2021socially}} & Roundabout &  Stackelberg Eq. & 2 & Continuous  \\ \hline
\makecell{\cite{stefansson2019human}} & Truck Platooning & Hierarchical Stackelberg Eq. &  2 &  Continuous ($a$ and $\delta$) \\ \hline
\makecell{\cite{yu2018human}} & Lane Change & Stackelberg Equilibrium & 2 & Discrete (Lane change) \\ \hline
\makecell{\cite{fisac2019hierarchical}} & Highway navigation & Hierarchical Feedback & 2+ & Continuous  \\ 
&&Stackelberg Eq. &&\\
\hline
\makecell{\cite{bahram2015game}} &  Highway navigation, &  Game Tree & 2 & Discrete \\ 
&Merge&&&\\
\hline
\makecell{\cite{isele2019interactive}} &  Merge &  Game Tree &     2  &    Discrete  \\ \hline
\makecell{\cite{laine2021multi}} &  Highway navigation & Generalised Feedback Nash Eq. & 2+ &   Continuous \\ \hline
\makecell{\cite{zhao2021yield}} & Left Turn & Generalized Nash Eq., SVO & 2 & Continuous ($a$ and $\delta$)  \\ \hline
\makecell{\cite{wang2020multi}} & Drone Racing & Generalized Nash Eq. & 6  &  Continuous ($v$)  \\ \hline
\makecell{ \cite{cleac2019algames}} & Merge, Intersection & Generalized Nash Eq. & 4 &  Continuous ($a$ and $\omega$) \\ \hline
\makecell{\cite{li2018game}} & Intersection & Level-k & 2 &   Continuous ($a$ and $\omega$) \\ \hline
\makecell{\cite{tian2018adaptive}}  & Roundabout & Level-k & 2 &  Continuous ($a$ and $\omega$)  \\ \hline
\end{tabular}
}
\end{table}

Game Theory examines equilibrium solutions under optimal player assumptions, with multiple concepts applicable to trajectory games. Dynamic games classify into open-loop and feedback games based on available information; open-loop assumes that the only information available to each player is the initial state of the game. For feedback games, the information available to each agent is the current state of the game.  Although the second type of game describes more accurately the AD setting,  Open-Loop solutions are often preferred for their simplicity. Common equilibria in Autonomous Driving include Open-Loop Nash, Open-Loop Stackelberg, Closed-Loop Nash, and Closed-Loop Stackelberg equilibria. For more details on the topic see \cite{bacsar1998dynamic}.

When the agents' dynamics have to comply with a set of constraints, such as collision-avoidance constraints, the equilibria are called \textit{generalized}. Generalized equilibria are studied in \cite{facchinei2010generalized}. Numerical solutions to the problems of Open Loop Nash Equilibria can be found in \cite{cleac2019algames, di2019newton, facchinei2009generalized}. The drawback of the Open-Loop Nash equilibrium formulation is that the players cannot reason directly on how their actions influence the behaviour of surrounding agents. A first simplification of this is the Open-loop Stackelberg equilibrium, which has been applied, for instance, in \cite{spica2020real} in the context of drone-autonomous racing. In a Stackelberg competition, the leader makes the first move, and subsequent players follow sequentially, allowing those with higher precedence to consider how those with lower precedence will plan their actions. 
In \cite{liniger2019noncooperative}, the authors propose a sequential bimatrix game approach to autonomous racing based on an Open-Loop Stackelberg game formulation. Other applications of Stackelberg formulation can be found in \cite{wang2020multi, hang2020integrated, yoo2020stackelberg}.
A formulation for solving generalized Feedback Nash equilibria can be found in \cite{laine2021computation}. Other methods can be found in \cite{fridovich2020efficient, fridovich2020iterative, schwarting2021stochastic}. 
Sadigh et al. \cite{sadigh2018planning} model AV-human interaction as a Partially Observable Stochastic Game in a Stackelberg competition. The human estimates the AV's plan and acts accordingly, while the AV optimizes its own actions, assuming indirect control over the human's actions. 

Typically, game-theoretic approaches suffer from the following problems: (1) the computational complexity grows exponentially in the number of agents and with the increasing temporal horizon, (2) they assume that the utility function that justifies other agents' actions is known to the ego-vehicle and that the agents act rationally with respect to those reward functions - however it is a known fact in game-theoretic finances problems that humans often do not act rationally \cite{sun2019interpretable}; (3) the behaviour of the agents might be stochastic and solving for mixed or behavioural strategies makes the computations even more intractable. 
Naturally, game theory also has the great advantage of capturing the interdependence of actions and some exact solutions exist for a restricted number of problems. Many papers in the field of game-theoretic autonomous driving try to alleviate those issues by further simplifying the problem or by finding approximate solutions. Now, a look will be taken at some papers in the field to analyze their simplifying hypotheses.

Level-k theories break down the Nash-equilibrium rational expectations logic by assuming people see others as being less sophisticated than themselves. This is level-k reasoning, in which the iteration process is stopped after k steps. Other agents are modelled as level k-1 actors. A level-k agent assumes that all the other agents are level-(k-1), makes predictions based on this assumption, and responds accordingly.
In \cite{tian2018adaptive}, level-k reasoning is applied to a roundabout scenario. This approach has also been incorporated in an RL framework in \cite{ding2018game}: the authors restrict the problem to two interacting agents and solve the Markov Game with two vehicles with a DQN RL-based approach.
In \cite{li2018game}, level-k reasoning is adopted to resolve conflicts at intersections. The authors showed that conflicts can be resolved easily in the case when the ego-vehicle is a level-k agent and all surrounding vehicles are level-k-1 or inferior. However, the number of collisions increases when both agents are of the same level, which indicates that further improvements need to be added to tackle scenarios with agents of the same kind, which would be crucial in the case of multiple AVs.

In order to keep computational complexity tractable, the number of agents can be reduced by identifying a subset of all agents that interact with the ego-vehicle \cite{stefansson2019human, isele2019interactive}. The time horizon can also be limited by considering a receding horizon controller or by implying hierarchical game-theoretic planning. The latter consists of having a short-horizon tactical planner in combination with a long-horizon strategic planner. The first one is responsible for accurate dynamics modelling of the problem, and the second one is responsible for deciding the strategy with approximate dynamics. Examples of this approach can be found in \cite{stefansson2019human, fisac2019hierarchical}.

Iterative linear-quadratic (LQ) methods are increasingly common in robotics and control communities. The authors of \cite{fridovich2020efficient} formulate the problem as a general-sum differential game characterized by nonlinear system dynamics. 
In \cite{fridovich2020iterative} extend their methods to systems with feedback-linearizable dynamics. 

Another method to solve game theoretic problems is to use Iterative Best Response to calculate Pure Nash Equilibria, i.e. Nash Equilibria in pure strategies.
The authors of \cite{wang2020multi} propose a ``sensitivity enhanced" iterative best response solver.
In \cite{wang2021game} an online game-theoretic trajectory planner based on the IBR is presented. The planner is suitable for online planning and exhibits complex behaviours in competitive racing scenarios. Williams et al. \cite{williams2018best} propose an IBR algorithm together with an information-theoretic planner for controlling two ground vehicles in close proximity.

In \cite{schwarting2019social}, Schwarting et al. propose an alternative method to the iterative best response to solve the Nash equilibrium problem based on a reformulation of the optimisation problem as a local single-level optimisation using the Karush–Kuhn–Tucker conditions. 
In \cite{zhao2021yield}, game theory is used to model other vehicles' decision-making. They propose a Parallel-Game Interaction Model (PGIM) to serve active and socially compliant driving interactions. 
To address uncertainty in the environment, \cite{schwarting2021stochastic} extend the game theoretic Nash equilibrium concept to POMDPs. In \cite{laine2021multi}, the authors take uncertainty about the intention of other agents by constructing multiple hypotheses about the objectives and constraints of other agents in the scene.

\section{Discussion and Future Challenges}
\label{sec:Discussion and Future Challenges}
In this comprehensive review, two critical sections central to the advancement of autonomous driving were introduced: \textit{Human Behaviour Studies} and \textit{Interaction Modelling}. These sections form the base for understanding and optimizing the intricate dynamics of interactions within autonomous driving scenarios. In this Section, challenges and research directions are highlighted for future autonomous vehicle research in interaction scenarios.

\subsection{Human Behaviour Studies}
Driven by society's strong desire for autonomous driving, human behaviour research has once again become a hot topic in recent years, especially research in AV contexts. In order to better understand pedestrian behaviour during AV interactions, many challenges still need to be overcome.

In general, the exploration of Driver Behavioral Models holds promise as a research area with the potential to bring about substantial enhancements in the safety and efficiency of transportation systems. Nevertheless, there is a substantial amount of work yet to be undertaken in the development and validation of these models. Future research should prioritise the creation of more holistic models that encompass a broader spectrum of factors, including the driver's psychological state, the surrounding environment, and interactions with other human participants on the road. 

For pedestrian behaviour studies, one critical challenge is communication. Firstly, although most researchers agree on the effectiveness of eHMIs,  a consensus is still lacking regarding their contents, forms, and perspectives \cite{bazilinskyy2019survey}.  An open question persists regarding whether eHMIs should be anthropomorphic or non-anthropomorphic. A similar question arises for textual and non-textual eHMIs. Furthermore, given the presence of multiple pedestrians on the road, the current eHMIs are primarily designed for one-to-one encounters, which may mislead other pedestrians \cite{dey2021towards}. Numerous comparable problems persist, which impede the standardisation of the eHMI. On the other hand, since implicit signals, such as vehicle kinematics,  are widely accepted, pervasive, common, and reliable communication methods, their critical roles cannot be ignored \cite{tabone2021vulnerable}. Although researchers have made initial attempts to influence pedestrians by manipulating implicit signals, such as vehicle deceleration rate, lateral distance, and pitch, these efforts are insufficient to ensure safe and efficient communication \cite{dey2021communicating,sripada2021automated,dietrich2020implicit}. These communication methods lack relevant theoretical support to demonstrate the accurate and effective delivery of communication information. Additionally, there is a dearth of reliable research paradigms to guide the research methodology, including vehicle driving behaviour design, subjective and objective experimental design, and more. Additionally, how to combine eHMI and implicit signals efficiently and smoothly to take advantage of both parties is also an interesting research direction.

Another challenge is pedestrian behaviour studies. Pedestrian decision-making and behavioural patterns are influenced by the diversity of interaction situations, traffic environments, and participants. However, these aspects currently lack sufficient research attention. Existing studies often focus on specific and simple interaction scenarios to control variables or simplify research complexity. However, real-life situations involve a plethora of complex scenarios, including crossings at multi-lane, two-way, or unstructured roads, crossings facing dense continuous traffic flow, multi-pedestrian crossing scenarios, and more. Moreover, pedestrian heterogeneity, such as gender, age, distraction, and group effects, also plays a significant role in interactions. It is noteworthy that many influential factors, such as waiting time and distraction, still lack a consensus. Accordingly, Due to the dearth of sufficient and reliable results, research conclusions mostly rely on assumptions, highlighting the inadequacy of understanding the underlying mechanisms of pedestrian road behaviour. 

Regarding pedestrian behaviour modelling, learning-based methods have been appealing in recent years. The end-to-end deep neural networks can effectively capture complex behavioural mechanisms and have made considerable progress in the field of pedestrian intention prediction and trajectory prediction. However, its \textit{black box} nature cannot be ignored. These approaches require a significant amount of data to achieve robust performance, which limits their scalability to sporadic cases with insufficient data. Additionally, the \textit{black box} models have difficulties in explaining their decision-making and behaviour logic, which brings new problems to modelling. In contrast, expert models, such as the Social Force models, Evidence Accumulation models, or Game Theoretical models, have solid psychological and behavioural foundations, and their behavioural decision-making logic is clear and interpretable. However, most of these models have only been validated on limited datasets or are still in the stage of laboratory validation, lacking extensive engineering practice. Hence, the theories of expert models need to be further refined and extensively verified on a large number of real datasets in the future. In addition, expert models and data-driven models have advantages in different aspects. A possible future trend is to find a balance point where the two models are used together.

Finally, considering that only a small fraction of the overall literature on autonomous driving explicitly considers the behaviour of pedestrians, there is a need to increase the application of pedestrian behaviour models, potentially including but not limited to pedestrian behaviour prediction, AV behaviour design, and virtual AV validation.

\subsection{Interaction Modelling}
As autonomous driving technology continues to evolve, research in interaction modelling will play a critical role in addressing challenges and driving the development of safer and more reliable autonomous vehicles. 

One prominent approach that has garnered attention in autonomous driving research is the use of learning-based methods. These methods offer the appeal of end-to-end solutions \cite{schwarting2018planning}, directly mapping sensory inputs and destination knowledge into the ego vehicle's actions. However, such systems can behave like a \textit{black-box}, leading to issues with interpretability in case of faults and the validation of their models. Furthermore, the enormity of the task, i.e., learning the entire driving process, poses a significant challenge. Hence, current research endeavours are breaking this down into sub-tasks, including route planning, perception, motion planning, and control and utilising learning-based methods to address these partial challenges.

The advantage of learning interactive behaviours through imitation learning or simulation in Deep Reinforcement Learning (DRL) approaches is also gaining momentum. Nevertheless, challenges persist. Most decision-making with Deep Learning assumes ideal road scenarios and a perfect perception of the surrounding environment. Yet, real-world conditions often involve occlusions, sensor noise, and environmental anomalies. Maintaining system performance in these sporadic events and handling partial or noisy information is an ongoing research challenge. Uncertainty seeps in from unpredictable behaviours in surrounding traffic participants, as well as sensor noise and vehicle models. Furthermore, models trained in simulated environments (like DRL models) raise the question of how to bridge the gap between simulation and reality. Several strategies have been proposed, including making simulations more realistic, domain randomisation, and domain adaptation \cite{tobin2017domain}. These approaches aim to prepare models for the unpredictability and complexity of the real world, ensuring that what they've learned in simulation can be applied effectively on the road. \\

An alternative approach to learning-based methods is model-based methods. This set of methods includes Game Theoretic models, Behavioural Models (which have been discussed in the previous Section), Social Forces and Potential Fields. 

Game theory offers versatility and adaptability and can effectively handle various scenarios without relying on specific data distributions. One of its key advantages is the ability to address planning and prediction for agents in a given situation. 
However, there's a trade-off in terms of computation. As the number of agents and the time horizon grow, the computational burden increases. Researchers have put forth several strategies to enhance game-theoretic solutions, including hierarchical game-theoretic formulations \cite{fisac2019hierarchical}, limiting the optimisation problem for surrounding agents to approximate solutions \cite{sadigh2018planning}, level-k game theory \cite{tian2018adaptive}, or improving nonlinear optimisation solvers' performances \cite{cleac2019algames, fridovich2020iterative, fridovich2020efficient}.

On the other hand, Social Force or Potential Fields methods offer a fast-computing solution. They can be used to predict surrounding agents' behaviour but also for ego-vehicle control. SFMs rely on simplified assumptions about human behaviour. They often treat pedestrians as particles or agents with fixed characteristics, neglecting the cognitive aspects of human decision-making, which can lead to unrealistic representations of complex and dynamic human behaviours. Future research directions for these methods include incorporating cognitive elements or contextual information, such as road rules and traffic signals. Exploring the integration of machine learning techniques to improve the adaptability and predictive power of SFMs can also be a possible future research direction. \\

The research predominantly concentrates on vehicle-vehicle interactions, which undoubtedly play a crucial role in autonomous driving. However, there is a pressing need to develop methods that address interactions with human road users, particularly pedestrians. As the realm of autonomous driving continues to evolve, the elucidation of theories and models that govern communication and interaction with diverse road users takes on an increasingly technical relevance, promising to propel safety and efficiency within autonomous driving scenarios.


\vspace{10mm}
\textbf{Acknowledgements} \par 
This project is supported in part by the EPSRC NorthFutures project (ref: EP/X031012/1) and the European Regional Development Fund. Luca Crosato and Kai Tian contributed equally to this work as co-first authors.
\medskip

%
\clearpage
\bibliographystyle{MSP}
\bibliography{bibliography}



\clearpage

\begin{figure}
  \includegraphics[width=40 mm,height=50 mm,clip,keepaspectratio]{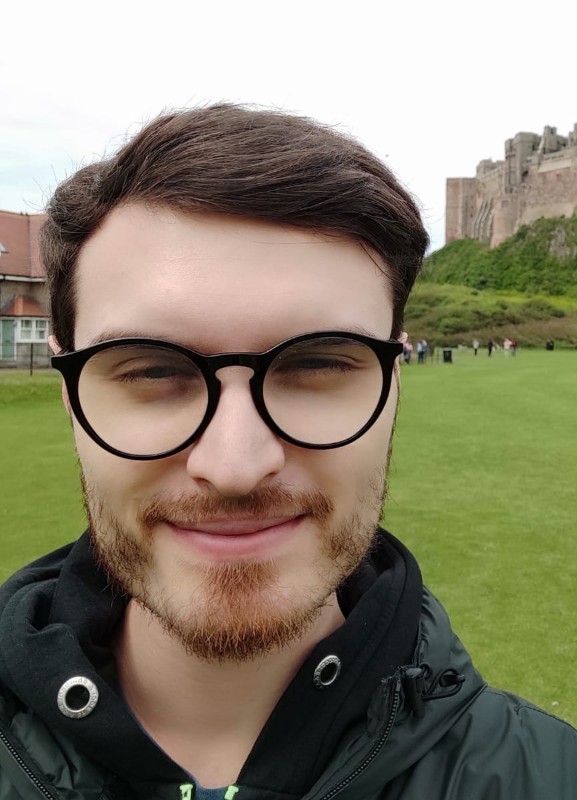}
  \caption*{Luca Crosato received a bachelor's degree in mechanical engineering and a master's degree in robotics and automation engineering from the University of Pisa, Pisa, Italy, in 2018. He is currently working toward a Ph.D. degree with Northumbria University, Newcastle upon Tyne, U.K. He is currently a Research Assistant at Queen's University Belfast, U.K. His research interests include decision-making and control of autonomous vehicles, pedestrian motion simulation, and machine learning.}
\end{figure}

\begin{figure}
  \includegraphics[width=40 mm,height=50 mm,clip,keepaspectratio]{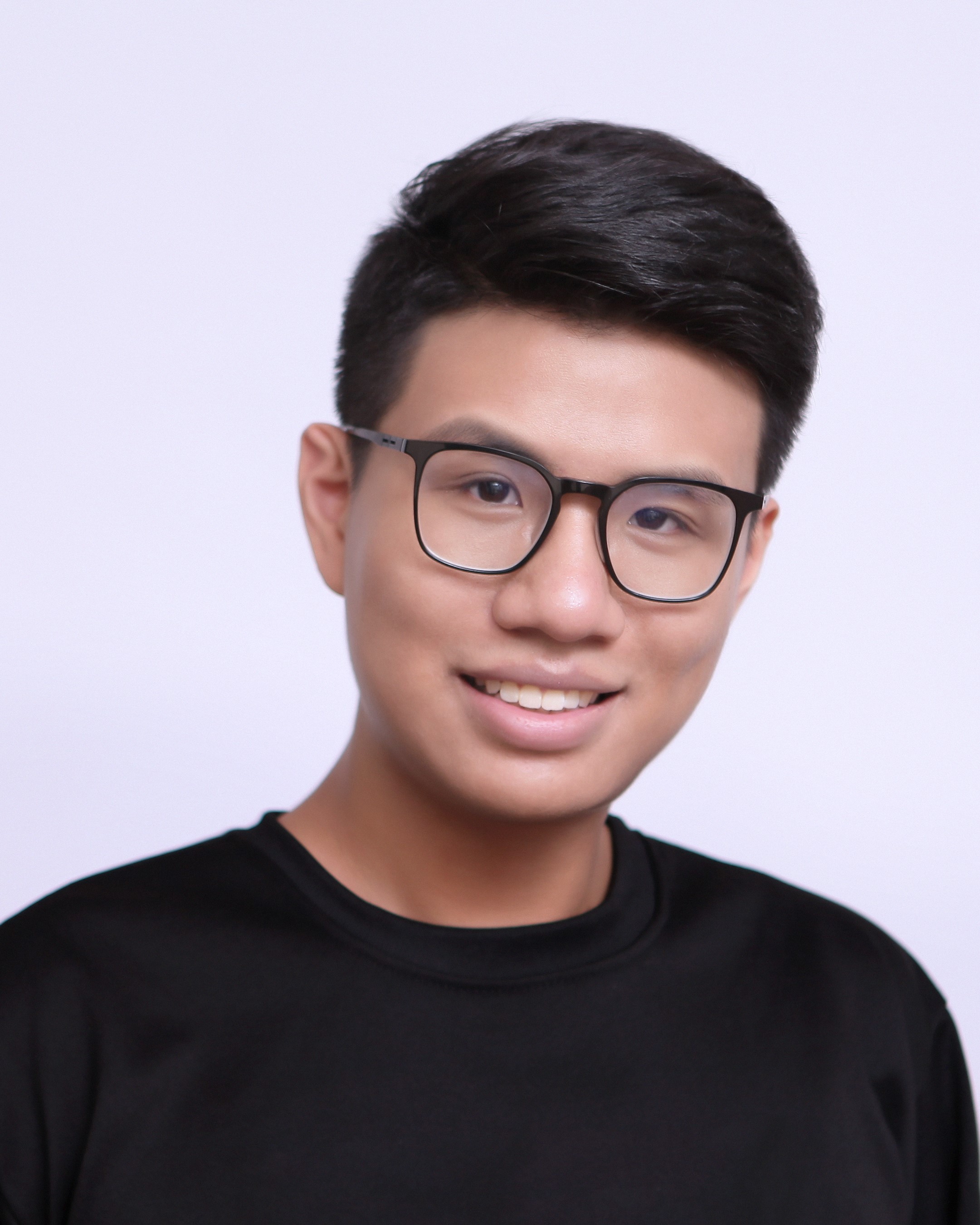}
  \caption*{Kai Tian received an M.Sc. degree in automotive engineering from Chongqing University, China, and a Ph.D. degree in transport Studies from the University of Leeds, UK. His main research interests include pedestrian-automated vehicle interaction, human factors and safety, and decision modelling.}
\end{figure}

\begin{figure}
  \includegraphics[width=40 mm,height=50 mm,clip,keepaspectratio]{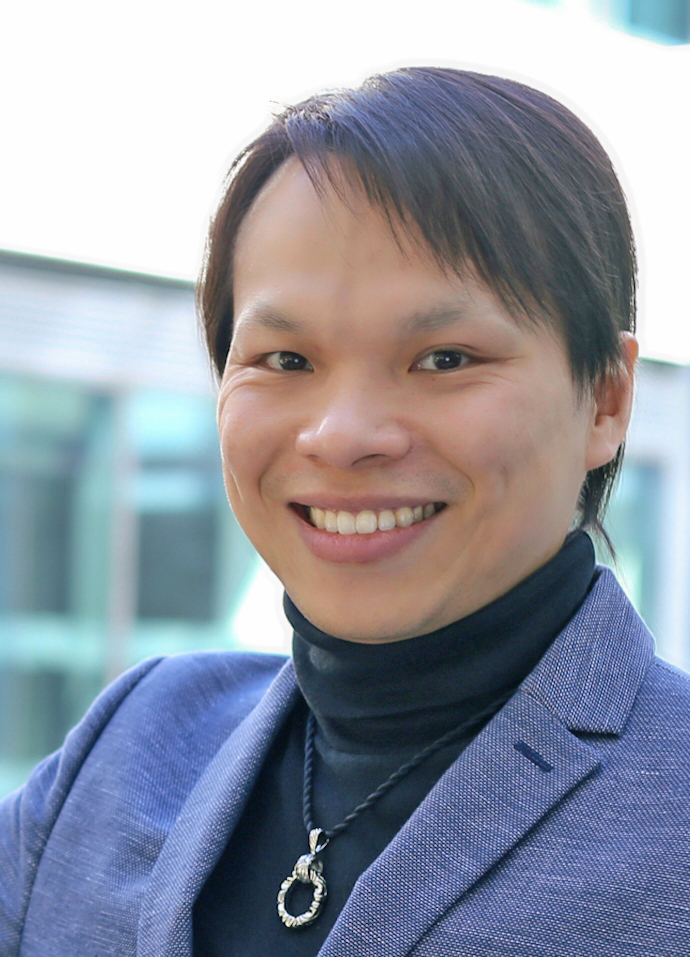}
  \caption*{Hubert P. H. Shum (Senior Member, IEEE) is an Associate Professor in Computer Science at Durham University, UK. Before this, he worked as the Director of Research/Associate Professor/Senior Lecturer at Northumbria University, UK, and a Postdoctoral Researcher at RIKEN, Japan. He received his Ph.D. degree from the University of Edinburgh, UK. His research interests include computer vision, computer graphics, motion analysis and machine learning.}
\end{figure}

\begin{figure}
  \includegraphics[width=40 mm,height=50 mm,clip,keepaspectratio]{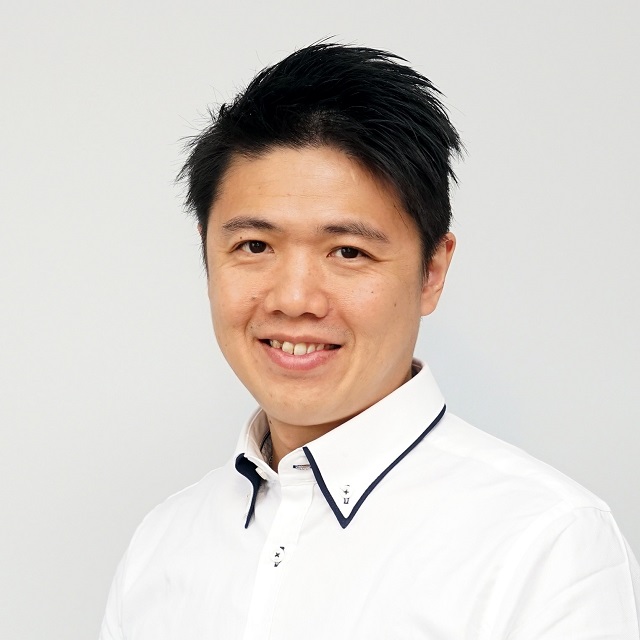}
  \caption*{Edmond Shu-lim Ho is currently a Senior Lecturer (Associate Professor) in the School of Computing Science (IDA-Section) at the University of Glasgow, Scotland, UK. Prior to joining the University of Glasgow in 2022, he was an Associate Professor in the Department of Computer and Information Sciences at Northumbria University, Newcastle upon Tyne, UK (2016-2022) and a Research Assistant Professor in the Department of Computer Science at Hong Kong Baptist University (2011-2016). He has been an Associate Editor of Computer Graphics Forum (CGF) since 2023. He received a BSc degree in Computer Science from the Hong Kong Baptist University, an MPhil degree from the City University of Hong Kong, and a PhD degree from the University of Edinburgh.}
\end{figure}

\begin{figure}
  \includegraphics[width=40 mm,height=50 mm,clip,keepaspectratio]{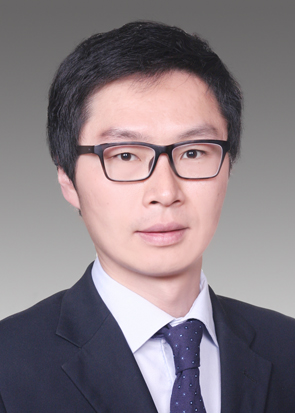}
  \caption*{Yafei Wang (Member, IEEE) received a B.S.degree in internal combustion engine from Jilin University, Changchun, China, in 2005, an M.S. degree in vehicle engineering from Shanghai Jiao Tong University, Shanghai, China, in 2008, and the Ph.D. degree in electrical engineering from The University of Tokyo, Tokyo, Japan, in 2013. From 2013 to 2016, he was a Post-Doctoral Researcher at The University of Tokyo. He is currently an Associate Professor at the School of Mechanical Engineering, Shanghai Jiao Tong University. His research interests include state estimation and control for connected and automated vehicles.}
\end{figure}

\begin{figure}
  \includegraphics[width=40 mm,height=50 mm,clip,keepaspectratio]{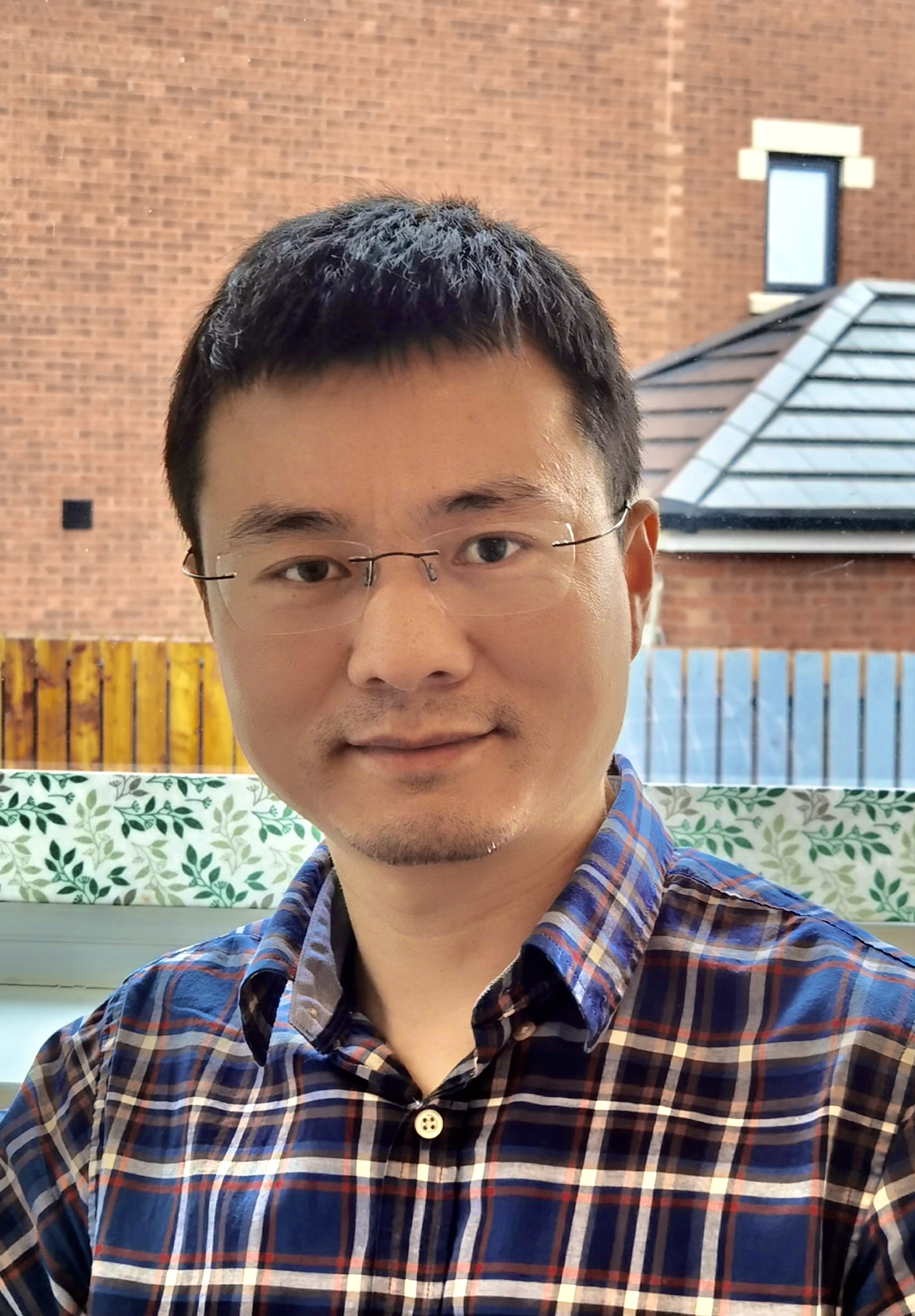}
  \caption*{Chongfeng Wei received a B.S. degree in computational and applied mathematics and an M.S. degree in vehicle engineering from Southwest Jiaotong University, Chengdu, China, in 2009 and 2011, respectively, and a Ph.D. degree in mechanical engineering from the University of Birmingham, Birmingham, U.K., in 2015. He is now a lecturer (assistant professor) at Queen's University Belfast, UK. His current research interests include decision-making and control of intelligent vehicles, human-centric autonomous driving, cooperative automation, and dynamics and control of mechanical systems. He is also serving as an Associate Editor of IEEE TITS, IEEE TIV, IEEE TVT, and Frontier on Robotics and AI.}
\end{figure}



\end{document}